\documentclass[twocolumn, switch]{article} 

\usepackage{preprint}

\usepackage{amsmath, amsthm, amssymb, amsfonts}

\usepackage[numbers,square]{natbib}
\usepackage[utf8]{inputenc}	
\usepackage[T1]{fontenc}	
\usepackage{xcolor}		
\usepackage[colorlinks = true,
            linkcolor = blue,
            urlcolor  = blue,
            citecolor = blue,
            anchorcolor = blue]{hyperref}	
\usepackage{booktabs} 		
\usepackage{nicefrac}		
\usepackage{microtype}		
\usepackage{lineno}		
\usepackage{float}			

\usepackage{lipsum}		

\usepackage{newfloat}
\DeclareFloatingEnvironment[name={Supplementary Figure}]{suppfigure}
\usepackage{sidecap}
\sidecaptionvpos{figure}{c}

\usepackage{titlesec}
\titlespacing\section{0pt}{12pt plus 3pt minus 3pt}{1pt plus 1pt minus 1pt}
\titlespacing\subsection{0pt}{10pt plus 3pt minus 3pt}{1pt plus 1pt minus 1pt}
\titlespacing\subsubsection{0pt}{8pt plus 3pt minus 3pt}{1pt plus 1pt minus 1pt}

\usepackage{tikz,xcolor,hyperref}

\definecolor{lime}{HTML}{A6CE39}
\DeclareRobustCommand{\orcidicon}{
	\begin{tikzpicture}
	\draw[lime, fill=lime] (0,0) 
	circle [radius=0.16] 
	node[white] {{\fontfamily{qag}\selectfont \tiny ID}};
	\draw[white, fill=white] (-0.0625,0.095) 
	circle [radius=0.007];
	\end{tikzpicture}
	\hspace{-2mm}
}
\foreach \x in {A, ..., Z}{\expandafter\xdef\csname orcid\x\endcsname{\noexpand\href{https://orcid.org/\csname orcidauthor\x\endcsname}
			{\noexpand\orcidicon}}
}

\title{Modeling of Hot-Carrier Degradation driven by silicon-hydrogen bond dissociation in SPADs}

\usepackage{authblk}

\author[1\thanks{\tt{mathieu.sicre@st.com}}]{Mathieu Sicre\orcidA{}}
\author[1]{Xavier Federspiel}
\author[1]{Bastien Mamdy\orcidB{}}
\author[1]{David Roy}
\author[2]{Francis Calmon\orcidC{}}

\affil[1]{STMicroelectronics, Crolles, France}
\affil[2]{INL, UMR, CNRS 5270, Université de Lyon, Villeurbanne, France}
\affil[*]{Email: mathieu.sicre@st.com}

\DeclareUnicodeCharacter{2212}{-}

\begin{document}

\twocolumn[ 
  \begin{@twocolumnfalse} 

\maketitle

\begin{abstract}
A novel approach for modeling  Dark Count Rate (DCR) drift ($\Delta$DCR) in Single-Photon Avalanche Diodes (SPADs) is proposed based on Hot-Carrier Degradation (HCD) inducing silicon-hydrogen bond dissociation at the Si/SiO$_{2}$ interface. The energy and the quantity of hot-carriers are modeled by the interplay of carrier energy distribution and current density. The carrier energy distribution, achieved by a Full-Band Monte-Carlo simulation considering the band structure and the scattering mechanisms, establishes a crucial link to the degradation of the top SPAD interface, primarily influenced by hot electrons due to their broader energy spread. The current density is determined by analyzing the generation rates of carriers under dark and photo conditions, along with the multiplication rate, through a combination of experimental data and modeling techniques. Subsequently, these hot carriers are correlated with the distribution of bond dissociation energy, which is modeled by the disorder-induced local variations among the Si-H bond energy at the Si/SiO$_{2}$ interface. The impact-ionization probability between hot carriers and Si-H bonds is then calculated by differentiating their energies, thereby determining the degradation kinetics. This enables the capture of the rise in dark current density with stress duration by the increasing number of defects, which in turn affects the modeling of degradation rate. For the first time, a direct correlation between the dark current and DCR, along with their drift over stress time, has been established, relying on the carrier generation rate originating from these defects together with the position-dependent breakdown probability P$_{\text{t}}$. This physic-based model allows to predict $\Delta$DCR for unprecedented long-term stress measurement time up to 10$^{6}$s, covering a whole set of characterization and stress conditions for SPAD devices.
\end{abstract}
\vspace{0.35cm}

  \end{@twocolumnfalse} 
] 



\section{\label{sec:introduction}Introduction}
\begin{figure}[h]
\centerline{\includegraphics[width=\columnwidth]{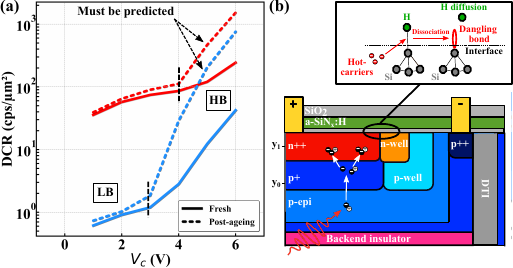}}
\caption{(a) DCR as a function of the characterization voltage V$_{\text{c}}$ measured at the characterization temperature T$_{\text{c}}$=300K (blue) and 333K (red). DCR is reported for pre-stress (solid lines) values and post-stress values (dashed lines) obtained at stress voltage V$_{\text{s}}$=5V, irradiance of 0.3W/m² at 940nm and temperature T$_{\text{s}}$=300K (blue) and 333K (red). (b) Cross section of the studied SPAD (not to scale). p-epi layer increases photon absorption. Highly doped n++ and p+ multiplication region confine avalanches at low biasing voltage. Lightly doped n and p guard rings prevent lateral breakdown at high biasing voltage.  y$_{0}$ and y$_{1}$ are the absorption depth at the surface and at the edge of the depleted volume  respectively. Inset: Zoom on the degradation localization with a schematic representation of hot-carrier degradation.}
\label{SPAD_cut_DCR_T_v1}
\end{figure}

Single-Photon Avalanche Diode (SPAD) Time-of-Flight (ToF) sensors are used for obstacle detection for automotive applications and gesture recognition for personal electronics \cite{9264255}. An emitter sends infrared light towards a scene and its reflection is detected by a SPAD receiver. The time difference between the emission and the reception provides the actual distance to the target. These photodetectors work by triggering an avalanche of charge carriers upon photon absorption, resulting in a substantial signal amplification which can be detected. Nevertheless, in low ambient light conditions, the maximum ranging distance of low reflectivity targets is limited by parasitically generated carriers triggering spurious avalanches called Dark Count Rate (DCR). An accurate physical description of the traps in term of parameters and localizations within the device is critical to accurately predict each individual DCR contribution \cite{10.1117/12.3001450,s23073412,Liu_2023,Dajing,8999742,Xu_2016}. Our previous studies \cite{SICRE1,SICRE2} shed light on the potential origins of the DCR thanks to voltage and temperature sweeps measured on non-stressed devices (bold lines in Fig. \ref{SPAD_cut_DCR_T_v1}\textcolor{blue}{a}). The low-bias region (LB) is dominated by carrier diffusion from interfaces and band-to-band tunneling triggering avalanches in the central n++/p+ region (Fig. \ref{SPAD_cut_DCR_T_v1}\textcolor{blue}{b}). The high-bias region (HB) is mainly driven by field-enhanced trap-assisted tunneling triggering avalanches in the peripheral n-/p-well region (Fig. \ref{SPAD_cut_DCR_T_v1}\textcolor{blue}{b}). 

In addition, the DCR drift ($\Delta$DCR) level must be predicted over time as a function of the operating conditions (dashed lines in Fig. \ref{SPAD_cut_DCR_T_v1}\textcolor{blue}{a}) to ensure signal integrity. Prior to our research, only one study had investigated the degradation of InP-based Geiger-mode avalanche photodiode arrays through the measurement of DCR over time under stress conditions and the analysis of current-voltage characteristics at different aging stages \cite{10.1117/12.819126}. Our previous work \cite{SICRE3} showed the susceptible degradation mechanisms responsible for $\Delta$DCR after stress. One of the culprits is charge accumulation at the top hydrogenated amorphous silicon nitride film (a-SiN$_{\text{x}}$:H) during temperature stress in darkness without any applied voltage. Another one is Hot Carrier Degradation (HCD) at the top depleted Si/SiO$_{2}$ interface that increases linearly with irradiance and exponentially with the inverse of the applied voltage related to the growth of carrier multiplications. The two degradation mechanisms are not correlated, and this paper will focus on building a model taking into account HCD mechanisms only. Our initial investigation \cite{SICREESS} primarily focused on examining the impact of carrier energy distribution on degradation. The ongoing study has broadened its scope to include not just carrier energy, but also the distribution of Si-H bond dissociation energy, along with the consideration of dark current drift.

In our previous works, $\Delta$DCR  was studied at a fixed stress duration to extract the stress condition dependencies and to identify degradation mechanisms in the SPAD architecture \cite{SICRE3}. The time dependence of $\Delta$DCR under device stress must be explained to reflect the physics of the interface defect formation process and to predict device reliability. Dangling bonds at the disordered Si/SiO$_{\text{2}}$ interface are electrically active and can capture/emit carriers. The role of these dangling bonds in the magnitude of the degradation was evidenced by the "giant isotope effect" \cite{doi:10.1063/1.116172}. Deuterium was used instead of hydrogen to anneal the dangling bonds at the Si/SiO$_{2}$ interface. The silicon-hydrogen/deuterium (Si-H/D) bonds at the interface break under hot carrier interactions and leads to hydrogen/deuterium diffusion (inset in Fig. \ref{SPAD_cut_DCR_T_v1}\textcolor{blue}{b}). This induces a change in the interface defect profile increasing the total net carrier generation. The bond dissociation energy of silicon-deuterium bonds is higher that of silicon-hydrogen bonds leading to lower bond-breakage rate at the interface. In the examined SPAD architecture, the transition from forming gas annealing (FGA) in a H$_{2}$ atmosphere to high-pressure deuterium annealing (HPD) during the annealing step resulted in a halving of $\Delta$DCR, confirming the occurrence of the giant isotope effect at the interface.

Before delving deeper, an overview of HCD in Metal-Oxide-Semiconductor Field-Effect Transistors (MOSFETs) will be synthesized \cite{TUW-237415}, initially focusing on carrier energy distribution and subsequently on Si-H bond energy distribution. This synthesis aims to support the proposed degradation model for SPAD devices.

The mechanisms behind HCD kinetics have been described by the Hess model \cite{1186774,a8e887eaa8c94af19390be1498d29f37,HESS19981} using both single- and multiple-carrier processes. Long-channel MOSFETs operated at high voltages, resulting in the presence of carriers with energies exceeding the bond dissociation reaction threshold. The large number of these energetic carriers makes it possible for a Single-Particle (SP) collision to initiate the bond dissociation process. In this case, HCD decreases as the temperature increases \cite{HONG1999809}. The scattering mechanism rates increase with temperature, and therefore depopulate the high-energy fraction of the carrier ensemble. As the device dimensions in short-channel MOSFETs decrease, the mean energy of the carrier energy distribution reduces and the bond dissociation can be initiated by a sequence of colder carriers that collide with the interface, inducing either phonon absorption or emission. The Si-H bond can be modeled as a quantum well with eigenstates that are filled by these bond excitation/deexcitation processes. When the energy acquired by the bond reaches the highest bonded level through Multiple Vibrational Excitation (MVE), the bond dissociation occurs. In the Rauch–La Rosa paradigm \cite{Childs1,Rauch2,Childs1996AOS,Rauch1998ImpactOE}, HCD is exacerbated at higher temperatures. The energy deposited by carriers drives HCD instead of the maximal electric field, which was in the lucky electron model \cite{Hu1979LuckyelectronMO}. On top of that, Electron-Electron scattering (EES) can populate the high-energy tail of the carrier ensemble strengthening the role of SP mechanism. Bravaix \textit{et al.} \cite{5173308,Bravaix2,Guerin2} summarize the three fundamental HCD modes which involve the SP mechanism, explained by the lucky electron model and marked by high average carrier energies, the MVE process, described by the truncated harmonic oscillator model for the Si-H bond, and linked to high carrier flux but low carrier energies, and the intermediate regime, governed by EES and characterized by moderate current densities and carrier energies.

To accurately model the degradation kinetics of hot carriers, it is necessary to consider the carrier energy distribution function, the Si-H bond distribution and the depassivation process. The Penzin model \cite{1213815} describes the bond rupture process by a kinetic equation for the passivated bond concentration with a depassivation and passivation reaction rate. The initial and broken bonds concentration is modulated by an attempt frequency and a hot-carrier acceleration factor. The activation energy of bond dissociation depends on the hydrogen density and the transversal component of the electric field. The Penzin model can also be derived in the form of a power law from the disorder-induced local variations among the Si-H bond energy at the Si/SiO$_{2}$ interface. The Bravaix model \cite{5173308,Bravaix2,Guerin2} integrates the interaction between the SP and MVE mechanisms found in the Hess model, and it replaces the carrier energy distribution function with operation/stress condition factors associated with knee energies in impact ionization and bond-breakage rates, following the Rauch–La Rosa approach. The Tyaginov model \cite{TYAGINOV20101267} introduces the carrier Acceleration Integral (AI) in the framework of Bravaix model which accounts for the carrier energy distribution function and the group velocity. The rate at which bonds are broken can be expressed as the product of the attempt frequency and the AI weighted by each mechanism's probability, where SP and MVE are the two mechanisms being considered.

In micrometer-large SPADs, the device operating voltages are high and carriers with energies above the threshold energy of the bond dissociation reaction E$_{\text{th}}$ are likely present in substantial density. Such energetic carriers can trigger the bond dissociation process in a single-particle collision. Hence, the MP mechanisms were disregarded for this reason. This assumption will be verified hereafter based on Full-Band Monte-Carlo simulations which compute the carrier distribution as a function of energy considering band structures and scattering mechanisms. The dark- and photo-generated current density coupled with the multiplication one will be used to determine the number of colliding hot carriers. The impact-ionization probability will be modeled by differentiating the carrier energy and the Si-H bond energy distributions. The generation rate of carriers from the dangling bond defects will be integrated with the avalanche breakdown probability P$_{\text{t}}$ to calculate $\Delta$DCR along stress time under different characterization and stress conditions. To summarize, the proposed model considers the concept of carrier AI accounting for the current density rather than the group velocity, in conjunction with the Si-H bond energy distribution exhaustion.

\section{\label{sec:Exp}Experimental and simulation framework}

The studied back-side illuminated SPADs are based on a silicon n-on-p structure \cite{Hofbauer,Moussy} (Fig. \ref{SPAD_cut_DCR_T_v1}\textcolor{blue}{b}) described in \cite{SICRE1,SICRE2} stacked by hybrid-bonding \cite{9732895} to an integrated Passive Quenching Circuit (PQC) fabricated in a standard 40nm CMOS technology \cite{Pellegrini}. When an avalanche is triggered in a SPAD, it generates a large current that results in a potential drop within the device that can be measured as an output signal pulse. However, to allow new event detection and avoid physical breakdown, the photodetector signal current must be limited and rapidly reduced. For this purpose, both active and passive quenching circuit techniques have been used \cite{Cova:96,Tisa2008VariableloadQC} to suppress the avalanche by lowering the reverse bias of the diode below the avalanche breakdown voltage. In a PQC, a resistor or a transistor is placed in series with the SPAD. The avalanche current discharges the junction and stray capacitances so that the voltage and current exponentially fall toward the asymptotic steady state values. The quenching time can be adjusted by changing the values of the resistor or capacitor used in the circuit.

\begin{figure}[h]
\centerline{\includegraphics[width=\columnwidth]{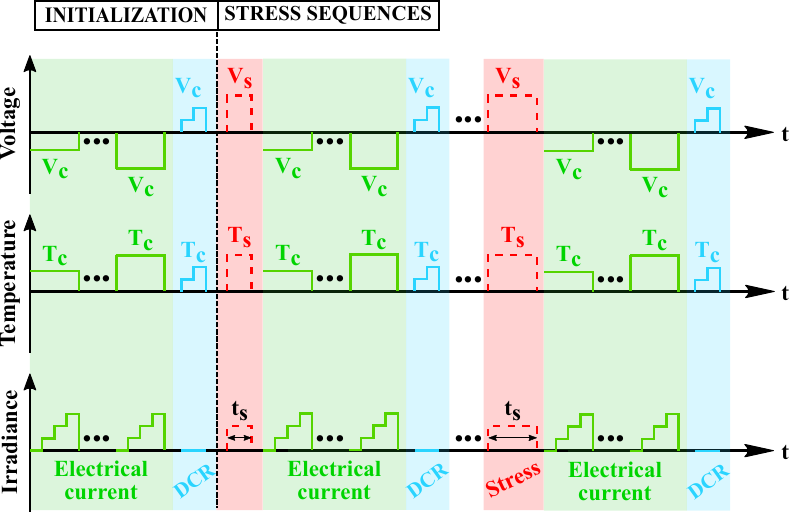}}
\caption{Time chronograph of initialization and stress sequences. Bold lines are measurement and dashed lines are stress. Electrical current is measured at different V$_{\text{c}}$,   T$_{\text{c}}$ and irradiance before stress (green box). DCR measurements are performed before stress and after each stress phase at different V$_{\text{c}}$ and T$_{\text{c}}$ (blue box). Devices are stressed (red box) for an increasing stress time t$_{\text{s}}$ at V$_{\text{s}}$, T$_{\text{s}}$ and irradiance.}
\label{DCR_pre_post_chrono_t_v2}
\end{figure}

Measurements and stresses are performed on eight SPADs stressed for each 24-hours stress condition reported in {{Table}} \ref{tab1}. Measurements focus on obtaining data to quantify device properties along device lifetime, while stress testing involves subjecting devices to aging conditions to evaluate their response and performance under stress. The electrical current I below the LB avalanche breakdown voltage V$_{\text{BD}}^{\text{LB}}$ is measured before and after the stress sequences (green color in Fig. \ref{DCR_pre_post_chrono_t_v2}) to extract the current density as a function of temperature and voltage. DCR is measured  with respect to temperature and voltage both before and after each stress period (blue color in Fig. \ref{DCR_pre_post_chrono_t_v2}). The absolute $\Delta$DCR corresponding to the difference between the DCR at a given stress time and the initial DCR is then computed. For comparison purposes, the DCR is normalized per unit photo-sensitive area (cps/µm²). DCR measurements are performed at characterization temperature T$_{\text{c}}$ and voltage V$_{\text{c}}$ applied above V$_{\text{BD}}^{\text{LB}}$ while the diode is in reverse bias condition. Stresses are performed either in darkness or under light at stress temperature T$_{\text{s}}$ and voltage V$_{\text{s}}$ applied above or below V$_{\text{BD}}^{\text{LB}}$ while the diode is in reverse bias condition. The extrinsic devices showing higher DCR levels, resulting likely from implant defects in the junction \cite{SICRE2}, are not considered to ensure these do not affect the average behavior of the population.

\begin{table}[h]
\footnotesize
\centering
\caption{Stress conditions}
\begin{tabular}{ c c c }
\toprule
\toprule
T$_{\text{s}}$ (K)& 
V$_{\text{s}}$ (V)& 
Irradiance (W/m²)\\

\hline 

333&
-3/-1/1/2/3/4/5 & 
0.0 \\

\hline

333& 
-9/-6/-3/-1/1/3/4/5 & 
0.3 \\

\hline

300/333/373& 
3 & 
0.0/0.3 \\

\hline

353& 
-3/4/5 & 
0.0\\

423& 
5 & 
0.0\\

353/373/423& 
-3 & 
0.33\\

353&
-3 & 
0.01\\

353	&	
4 & 
0.1/0.33/0.43\\

353	&	
5 & 
0.08/0.33/0.42\\

373	&	
5 & 
0.22\\

\bottomrule
\bottomrule

\end{tabular}
\label{tab1}
\end{table}

\begin{figure}[h]
\centerline{\includegraphics[width=\columnwidth]{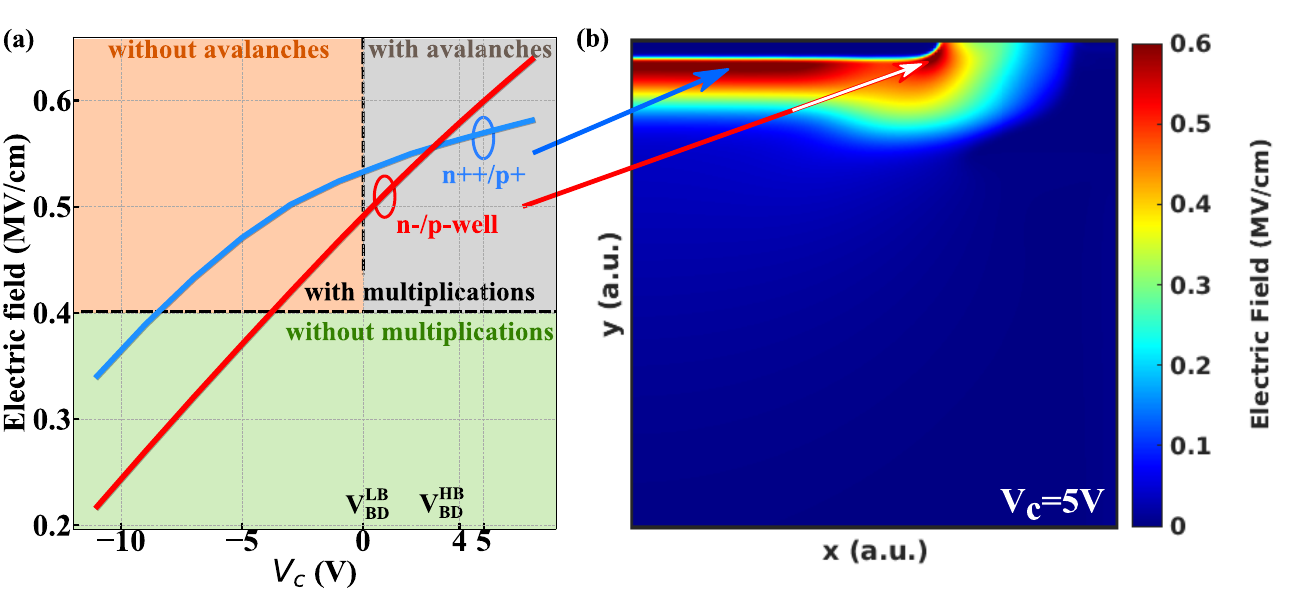}}
\caption{(a) Electric field as a function of V$_{\text{c}}$ normalized by the central avalanche breakdown voltage V$_{\text{BD}}^{\text{LB}}$ in n++/p+ (blue) and n-/p-well (red) region simulated at T$_{\text{c}}$=333K. Depending on the electric field magnitude, there is or not multiplication either in n++/p+ or n-/p-well region (green and orange box) without self-sustaining avalanches. Above V$_{\text{BD}}^{\text{LB}}$, there are  avalanches in the n++/p+ region (gray box). Avalanches in the n-/p-well are triggered at higher voltage biasing V$_{\text{BD}}^{\text{HB}}$ which is the peripheral avalanche breakdown voltage. (b) Electric field map simulated at V$_{\text{c}}$=5V and T$_{\text{c}}$=333K underlining the electric field activity in both region.}
\label{E_HB_v1}
\end{figure}

\begin{figure}[h]
\centerline{\includegraphics[width=\columnwidth]{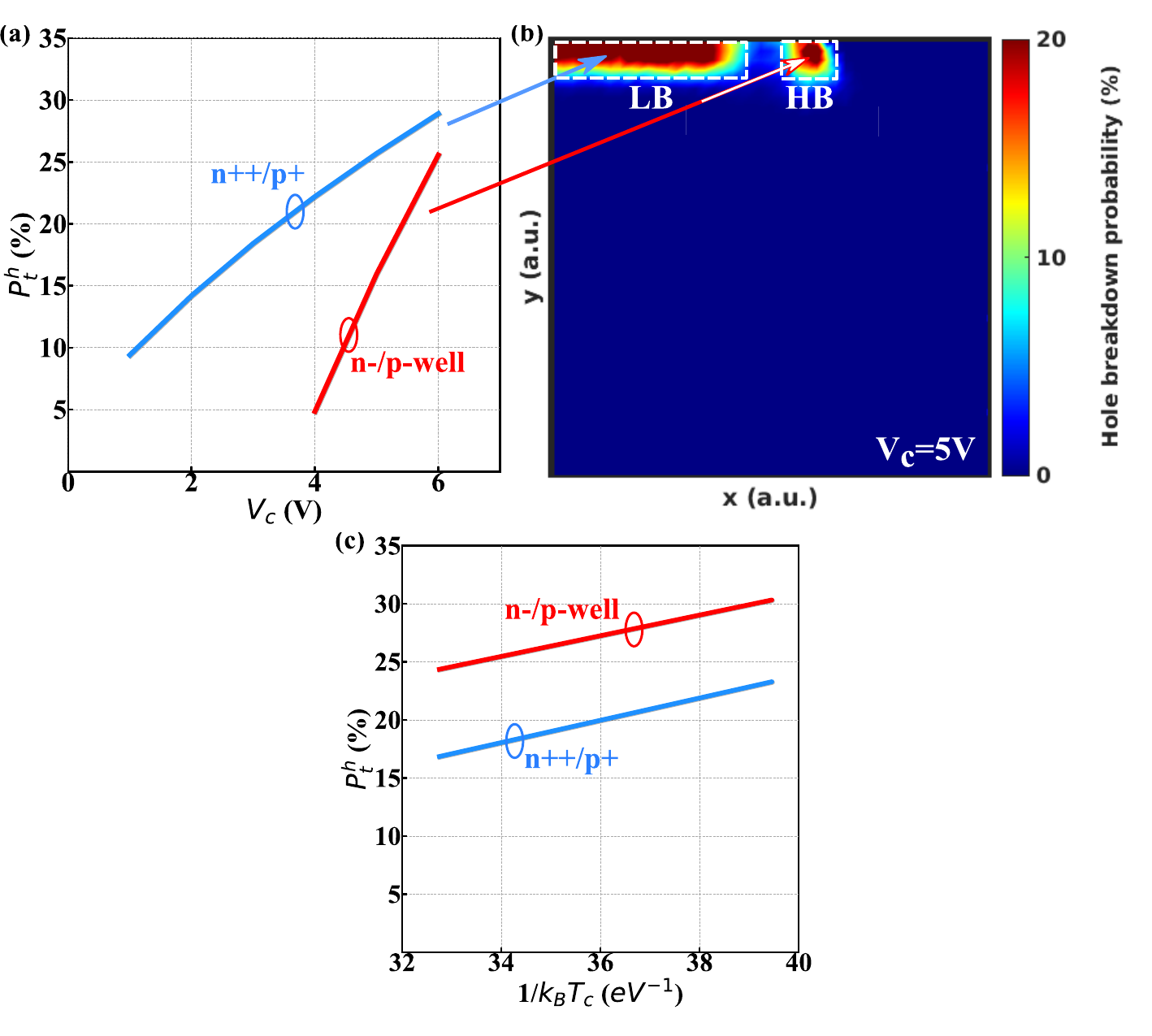}}
\caption{ (a) Hole breakdown probability computed with a drift-diffusion Monte-Carlo \cite{SICRE2,Jacoboni_2} as a function of V$_{\text{c}}$ in n++/p+ (blue) and n-/p-well (red) region simulated at T$_{\text{c}}$=333K. The probability of hole breakdown refers to the likelihood of a hole initiating a self-sustaining avalanche. (b) Hole breakdown probability map simulated at  V$_{\text{c}}$=5V and T$_{\text{c}}$=333K. The n++/p+ region is responsible of LB dynamics and the n-/p-well region of HB one. (c) Hole breakdown probability as a function of 1/k$_{\text{B}}$T$_{\text{c}}$ simulated in n++/p+ region at V$_{\text{c}}$=3V and in n-/p-well region at V$_{\text{c}}$=6V. k$_{\text{B}}$ is the Boltzmann constant.}
\label{Pt_TV_v1}
\end{figure}

The device structure was built by simulating fabrication processes through the Sentaurus Process tool \cite{sprocess}. The electrical outputs were computed by solving the Poisson and current continuity equations in the Sentaurus Device module \cite{sdevice}, consistent with the drift-diffusion transport model. The electric field dynamics against voltage in the n++/p+ and n-/p-well regions are shown in Fig. \ref{E_HB_v1}\textcolor{blue}{a}. Carrier multiplications are expected to rise above a certain electric field threshold \cite{PhysRev.109.1537}. A Monte-Carlo model \cite{Helleboid_2022, SICRE2,Jacoboni_2} accounting for the stochastic formulation of the drift-diffusion equation and the impact-ionization process was used to simulate P$_{\text{t}}$. There are two avalanche triggering dynamics against voltage in n++/p+ (LB) and n-/p-well (HB) region (Fig. \ref{Pt_TV_v1}\textcolor{blue}{a}). The positive thermal dependence of P$_{\text{t}}$ (Fig. \ref{Pt_TV_v1}\textcolor{blue}{c}) is a consequence of carrier multiplication \cite{PMARS}. The electron-phonon mean free path decreases with increasing lattice temperature which results in the positive shift of multiplication threshold for the same applied voltage at different temperatures. The integration of P$_{\text{t}}$ and the carrier generation rate from mid-gap neutral traps yields the DCR at each defect site. This approach will be employed to calculate the $\Delta$DCR arising from stress-induced defects.

\section{Current-induced degradation}

\begin{figure}[h]
\centerline{\includegraphics[width=\columnwidth]{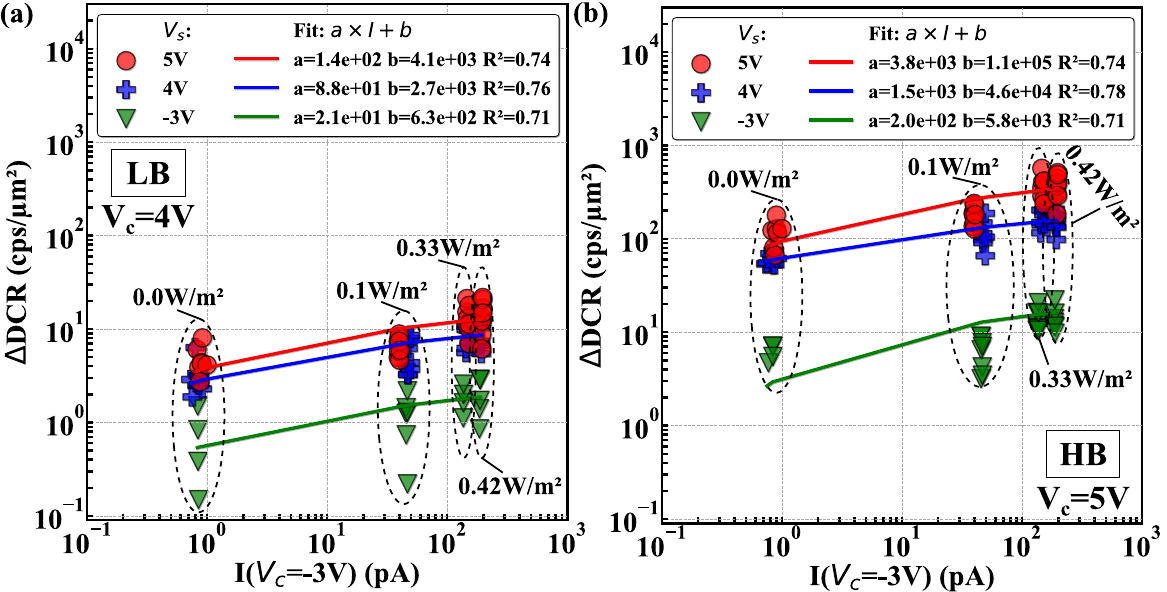}}
\caption{$\Delta$DCR as a function of the electrical current I (without avalanches) measured (symbols) and fitted (lines) by a linear relationship at V$_{\text{c}}$=-3V and at T$_{\text{c}}$=353K for several irradiance levels from 0.0W/m² (left) to 0.42W/m² (right). $\Delta$DCR was obtained from 24-hours stresses at different V$_{\text{s}}$ and at T$_{\text{s}}$ =353K for the same irradiance levels and was measured at T$_{\text{c}}$ =333K and at (a) V$_{\text{c}}$ =4V (LB) and (b) 5V (HB). }
\label{DDCR_IV_v1}
\end{figure}

Our previous study \cite{SICRE3} has shown that DCR degradation could occur even without avalanches. Avalanches may result in large amounts of hot-carriers but they cannot flow near the interface due to electrostatic potential drop \cite{Ingargiola}. A typical I-V curve above V$_{\text{BD}}^{\text{LB}}$ can be found in \cite{SICRE3}. When the diode is biased above V$_{\text{BD}}^{\text{LB}}$, it is impossible to differentiate between the dark- and photo-generated current, which do not trigger avalanches, and the avalanche current. However, it can be assumed that it is proportional to the current measured below V$_{\text{BD}}^{\text{LB}}$, in darkness or under illumination, accounting for the multiplication rate. The $\Delta$DCR after stress is therefore compared to the current measured below V$_{\text{BD}}^{\text{LB}}$ before stress and reported in Fig. \ref{DDCR_IV_v1} for both characterization regions. With (V$_{\text{s}}$=4V and 5V) or without (V$_{\text{s}}$=-3V) avalanche triggering, $\Delta$DCR versus current measured without avalanches follows a linear relationship with the irradiance. The same aging mechanisms can be assumed to relate to the increasing number of hot carriers that enhance Si-H bond dissociation. According to these results, the current density without avalanches will be extrapolated above V$_{\text{BD}}^{\text{LB}}$ to compute the degradation rate.

\section{\label{sec:EC}Carrier energy distribution simulation}

A full-band Monte-Carlo method was used to demonstrate SP degradation process in SPADs through computation of carrier energy distribution n(E). All FBMC simulations have been performed with an in-house full band ensemble Monte Carlo solver (UTOX) \cite{MICHAILLAT20102437,RevModPhys.55.645}. The full-band Monte-Carlo approach stands on silicon band structures computed with Tight-Binding method \cite{Niquet}. However, the full-band Monte-Carlo method is time-consuming and demands significant computational resources. Instead, simplified approaches like Drift Diffusion (DD) and Energy Transport (ET) schemes \cite{GrasserJungemann,ZakaRafhay} for solving the Boltzmann transport equation (BTE) can be used to substitute the Monte-Carlo method of carrier transport treatment. The full-band Monte-Carlo approach stands on silicon band structures computed with Tight-Binding method. Phonon scattering, ionized impurity scattering, and carrier-carrier scattering rates are calculated consistently with the band structure. During the initialization stage, a rejection sampling in carrier energy density, which is the occupation probability at a given energy multiplied by the state density at this energy, is performed to determine the energy of each carrier. The wave vector $\overrightarrow{k}$ of each carrier is randomized on the resultant energy iso-surface at each time step. The carrier acceleration is computed according to the integration of the second Newton's law which leads to:

\begin{equation} \overrightarrow{\Delta k} = \frac{1}{\hbar}q\overrightarrow{E} \Delta t_{ff} \label{eq}\end{equation}

with $\text{$\hbar$}$ the reduced Planck's constant, q the elementary charge and $\overrightarrow{\text{E}}$ the electric field reported in Fig. \ref{E_HB_v1}\textcolor{blue}{a}. The carrier free-flight time $\Delta \text{t}_{\text{ff}}$ is computed by a rejection sampling in the interaction probabilities with the different scattering mechanisms. The scattering is either elastic (carrier-impurity) saving carrier energy, lowly inelastic (carrier-phonon or carrier-carrier) or highly inelastic (impact ionization or Auger recombination) leading to meV-exchange or to eV-exchange respectively. The impact ionization rate is modeled by the Keldysh formula $P_{II}=P_{II}^{0}(E-E_{0})^{\alpha_{0}}$ \cite{Keldysh1965ConcerningTT} with E and E$_{0}$ the carrier energy and the threshold impact-ionization energy, P$_{II}^{0}$ is the attempt rate and $\alpha_{0}$ the energy exponent of the impact-ionization dynamics \cite{7605145,10.1063/1.362375}. To simulate the post-ionization state, it was assumed that the carrier kinetic energy lost by the incoming particle during the impact is constant and equal to the band gap E$_{\text{g}}$ \cite{PhysRevB.38.9721,YAMADA1995881}. Exploring the impact of carrier energy distribution functions, both with and without carrier-carrier scattering, falls outside the scope of this paper.

\begin{figure}[h]
\centerline{\includegraphics[width=\columnwidth]{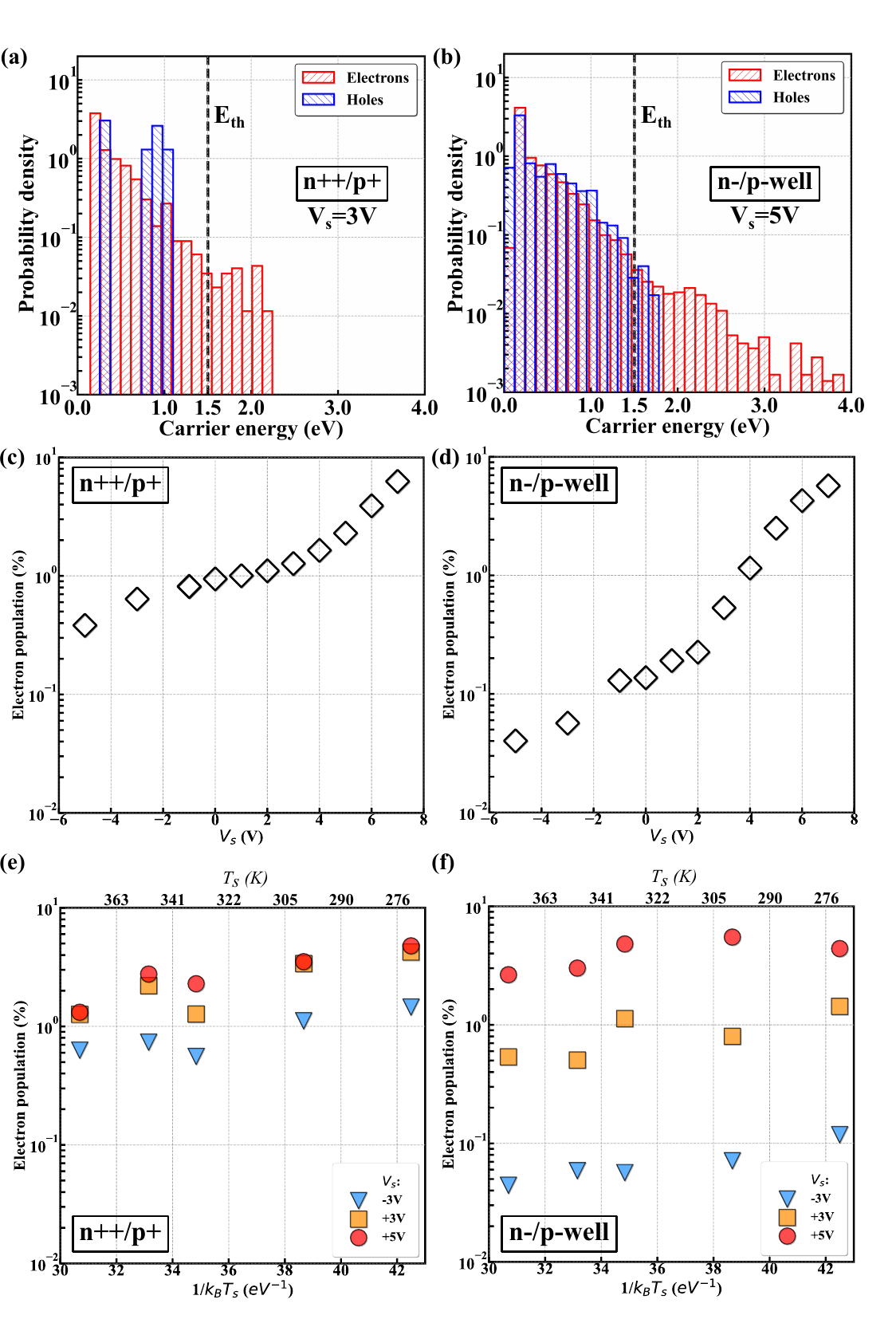}}
\caption{Probability density of electron (red) and hole (blue) energy simulated by a full-band Monte-Carlo \cite{MICHAILLAT20102437,RevModPhys.55.645} at T$_{\text{s}}$=333K for (a) V$_{\text{s}}$=3V in n++/p+ region (LB) and (b) V$_{\text{s}}$=5V in n-/p-well region (HB). Carriers acquiring energy above Si-H bond dissociation (1.5eV) are extracted and reported in (c) and (d) as a function of V$_{\text{s}}$ at T$_{\text{s}}$=333K and in {{(e)}} and {{(f)}} as a function of 1/k$_{\text{B}}$T$_{\text{s}}$ at V$_{\text{s}}$=-3V (blue triangles), 3V (orange squares), 5V (red circles).}
\label{Carrier_distrib_v1}
\end{figure}

The top interface was previously identified as degradation localization in the SPAD architecture \cite{SICRE3} thus a focus on n(E) in its vicinity is realized. At the beginning of the simulation, one hundred electron-hole pairs are placed in either the vicinity of the n++/p+ or n-/p-well region. The electrons move along the upper interface toward the cathode (not shown here), while the holes travel in opposite direction as previously demonstrated with the drift-diffusion Monte-Carlo simulations \cite{SICRE2,SICRE3}. n(E) is extracted at the top interface either in n++/p+ or n-/p-well region and normalized by the total carrier distribution obtained in the simulation (Fig. \ref{Carrier_distrib_v1}\textcolor{blue}{a-b}) to compare hot carrier populations between each condition. Some energy-tail electrons reach the mean threshold dissociation energy of Si-H bond (E$_{\text{th}}$=1.5eV \cite{5173308,PhysRevB.59.12884}) depending on V$_{\text{s}}$ (Fig. \ref{Carrier_distrib_v1}\textcolor{blue}{c-d}) and T$_{\text{s}}$ (Fig. \ref{Carrier_distrib_v1}\textcolor{blue}{e-f}). This value corresponds to the bending vibrational mode, while recent ab initio calculations published in \cite{PhysRevB.100.195302} suggest that the bond dissociation reaction may occur via the stretching mode with a higher activation energy. The positive thermal dependence of carrier energy is a consequence of carrier multiplication \cite{PMARS} as previously explain for P$_{\text{t}}$ in {{Section}} \ref{sec:Exp}. The same thermal and voltage dependencies have been observed on the overall population. There are two potential degradation dynamics according to the localization either in n++/p+ or n-/p-well region. These two different degradation regions can be stimulated independently depending on the characterization conditions (Fig. \ref{Pt_TV_v1}). The simulated carrier energy distributions n(E) along V$_{\text{s}}$ and T$_{\text{s}}$ will be integrated over the bond dissociation energy to account for the bond-dissociation probability by hot carrier interaction at each carrier energy. 

Most of the equations presented are conventional within the field. Newly introduced and modified equations will be marked with an asterisk (*) placed at the right-hand side of the equation.

\section{Electric current modeling} 

\subsection{Multiplication current modeling}

Now that the carrier energy distribution is simulated for a given number of electron-hole pairs, the total carrier density flowing in the device must be known to model the degradation rate. As previously explained, the current density  above V$_{\text{BD}}^{\text{LB}}$ is modeled by extrapolating the current density below V$_{\text{BD}}^{\text{LB}}$ to get rid of the avalanche current. The multiplication rate is equal to:

\begin{equation}
G_{\alpha} = \alpha _{n} \frac{j_{n}^{d,ph}}{q} + \alpha _{p} \frac{j_{p}^{d,ph}}{q}
\label{G_a}
\end{equation}

j$_{\text{n,p}}^{\text{d,ph}}$ is the dark- and photo-generated current density. $\alpha _{\text{n}}$ and $\alpha _{\text{p}}$ are the ionization rates which are defined as the mean number of charges generated by distance unit traveled by an electron or a hole. The impact-ionization coefficient are defined according to the Chynotew law \cite{PhysRev.109.1537} which states that:

\begin{equation}
\left\{\begin{matrix}
\alpha _{n}=\gamma a_{n} exp(\frac{-\gamma b_{n}}{E}) \\ 
\alpha _{p} =\gamma a_{p} exp(\frac{-\gamma b_{p}}{E})
\end{matrix}\right.
\label{eq}\end{equation}

\begin{equation}
\gamma = \frac{\Lambda_{n,p} (T_0)}{\Lambda_{n,p}(T)} = \frac{tanh(\frac{\hbar \omega_{op}}{2k_{B}T_{0}})}{tanh(\frac{\hbar \omega_{op}}{2k_{B}T})}
\label{gam}\end{equation}

$\Lambda_{\text{n,p}}$ is the mean free path for the electrons and holes with respect to optical phonons. The coefficients  $\text{a}_{\text{n,p}}$, $\text{b}_{\text{n,p}}$, and $\hbar \omega_{\text{op}}$ were measured by van Overstraeten and de Man \cite{VANOVERSTRAETEN1970583}. $\hbar \omega_{\text{op}}$ is defined as the optical-phonon energy. $\text{T}_{\text{0}}$ and T are the reference and the absolute temperature, and $\omega_{\text{op}}$ is the angular frequency of optical phonons. $\gamma$ and  $\hbar \omega_{\text{op}}$ express the temperature dependence of the phonon gas against which carriers are accelerated. The multiplication current is equal to G$_{\alpha}$ multiplied by q and by the multiplication-sensitive area which is defined as the space charge region S$_{\text{SCR}}$. The multiplication current contribution is coupled with the dark- and photo-generated ones considering the simulated electric field E (Fig. \ref{E_HB_v1}\textcolor{blue}{a}). In both the dark current and photo-generated current equations, the second term denotes the influence of carrier multiplication. 

\begin{table}[h]
\footnotesize
\centering
\caption{Model parameters}
\begin{tabular}{c c c c}
\toprule
\toprule
Quantity & 
Units & 
Value &
Ref./Cmt.
\\
\hline

a$_{\text{n}}$  & 
cm$^{\text{-1}}$ & 
7.03 $\cdot 10^{5}$ &

\\

a$_{\text{p}}$ & 
cm$^{\text{-1}}$ & 
1.582 $\cdot 10^{6}$ $^{\mathrm{a}}$  / 6.71 $\cdot 10^{5}$ $^{\mathrm{b}}$ &

\\

b$_{\text{n}}$  & 
V$\cdot \text{cm}^{\text{-1}}$ & 
1.231 $\cdot 10^{6}$ &
\scriptsize \cite{VANOVERSTRAETEN1970583}
\\

b$_{\text{p}}$  & 
V$\cdot \text{cm}^{\text{-1}}$ & 
2.036 $\cdot 10^{6}$ $^{\mathrm{a}}$ / 1.693 $\cdot 10^{6}$ $^{\mathrm{b}}$ &

\\

T$_{\text{0}}$ & 
K & 
300 &

\\

$\hbar \omega_{\text{op}}$& 
eV & 
0.063 &

\\

\hline

N$_{\text{t}}^{\text{0}}$& 
cm$^{\text{-2}}$ & 
$10^{12}$ &
\scriptsize \cite{Nguyen}
\\

$\sigma_{\text{t}^{\text{0}}}$& 
cm$^{-1}$ & 
$10^{-15}$ &
\scriptsize \cite{SICRE1}
\\

E$_{\text{t}}$& 
eV & 
E$_{\text{g}}$/2 &

\\

\hline

m$_{0}$ $^{\mathrm{c}}$& 
kg & 
9.11 $\cdot 10^{-31}$ &

\\

$\frac{\text{m}_{\text{{n}}}}{\text{m}_{0}}$& 
1 & 
0.2  $^{\mathrm{d}}$ &
\scriptsize \cite{Zeghbroeck}
\\

$\frac{\text{m}_{\text{p}}}{\text{m}_{0}}$ & 
1 & 
0.49 $^{\mathrm{e}}$ &

\\

\hline

E$_{\text{g}}^{0}$& 
eV & 
4.73 $\cdot 10^{-4}$ &

\\

$\alpha_{\text{g}}$& 
eV$\cdot \text{K}^{-1}$& 
4.73 $\cdot 10^{-4}$ &
\scriptsize \cite{VARSHNI1967149}
\\

$\beta_{\text{g}}$& 
K & 
636 &

\\

\hline

E$_{\text{p}}$ & 
eV & 
0.015 &
\scriptsize \cite{Pankove_1972}
\\

\hline

N$_{\text{hb}}^{\text{0}}$& 
cm$^{\text{-2}}$ & 
$3 \cdot 10^{12}$ &
\scriptsize \cite{Lenahan}
\\

\hline

E$_{\text{th}}$& 
eV & 
1.5 &
\scriptsize \cite{5173308,903763,PhysRevB.59.12884}
\\

$\sigma_{\text{bd}}$& 
eV & 
0.1 &
\scriptsize \cite{Haggag}
\\

\hline

$\sigma_{\text{II}}^{0}$& 
eV$^{-\rho_{0}} \cdot$ cm$^2$  & 
$10^{-20}$ &
Arb. set
\\

$\rho_{0}$& 
1 & 
11 &
\scriptsize \cite{Haggag}
\\

\bottomrule
\bottomrule

\multicolumn{4}{l}{$^{\mathrm{a}}$ electric field range: 0.175-0.4 MV/cm}\\
\multicolumn{4}{l}{$^{\mathrm{b}}$ electric field range: 0.4-0.6 MV/cm}\\
\multicolumn{4}{l}{$^{\mathrm{c}}$ m$_{0}$ is the free electron rest mass}\\
\multicolumn{4}{l}{$^{\mathrm{d}}$ considering conduction band minimum at k=0}\\
\multicolumn{4}{l}{$^{\mathrm{e}}$ considering heavy hole valence band maximum at E=k=0}\\
\end{tabular}
\label{tab2}
\end{table}

\subsection{Dark current modeling}

The dark current is modeled using Shockley–Read–Hall recombination-generation rate of carriers \cite{PhysRev.87.835} according to previous results \cite{SICRE1,SICRE2} that showed presence of defects at the top depleted interface (where n$_{\text{i}}^{\text{2}}$ is much greater than n and p). Under the assumption of a worst-case scenario, considering an identical defect profile for capturing electrons and holes (with a matching capture/emission cross-section $\sigma_{\text{t}}$) and the amphoteric nature of defects \cite{BIEGELSEN1985879} (characterized by a single initial interface defect density N$_{\text{t}}^{\text{0}}$) defined at a singular mid-gap energy level E$_{\text{t}}$, the generation rate can be simplified to:

\begin{equation}
G_{d}=\frac{N_{t}^{0} n_{i} \sigma_{t} v_{th}^{n} v_{th}^{p}}{v_{th}^{n} exp(\frac{E_{t}-E_{F}^{i}}{k_{B}T}) + v_{th}^{p} exp(\frac{E_{F}^{i}-E_{t}}{k_{B}T})}
\label{Gd}\end{equation}

where $\text{E}_{\text{i}}$ is the intrinsic Fermi level, and $\text{E}_{\text{v}}$  and $\text{E}_{\text{c}}$  are valence and conduction band energy. The intrinsic density $\text{n}_{\text{i}}$ follows:

\begin{equation}
n_{i}=\sqrt{N_{c}N_{v}}exp(\frac{-E_{g}}{2k_{B}T})
\label{n_i}\end{equation}

where the effective densities of states in the conduction band $\text{N}_{\text{c}}$ and in the valence band $\text{N}_{\text{v}}$ are:

\begin{equation}
N_{c,v}=2(\frac{2 \pi m_{n,p} k_{B} T}{h^{2}})^{\frac{3}{2}}
\label{eq}\end{equation}

with $\text{m}_{\text{n,p}}$ the effective mass of electron and hole and h the Planck's constant. The band gap follows \cite{VARSHNI1967149}:

\begin{equation}
E_{g}=E_{g}^{0}-\frac{\alpha_{g} \cdot T^{2}}{\beta_{g} + T}
\label{eq}\end{equation}

where $\text{E}_{\text{g}}^{\text{0}}$ is the bandgap at 0K and $\alpha_{\text{g}}$ and $\beta_{\text{g}}$ some experimental parameters  depending on the semiconductor material. The Fermi level for an intrinsic semiconductor $\text{E}_{\text{i}}$ is:

\begin{equation}
E_{i}=\frac{E_{g}}{2}+\frac{3k_{B}T}{4}ln(\frac{m_{p}}{m_{n}})
\label{E_i}\end{equation}

The thermal velocity v$_{\text{th}}$ is equal to:

\begin{equation}{v_{th}}=\sqrt{\frac{3k_{B}T}{m_{n,p}}}\label{v_th}\end{equation}

The electric field-enhancement of generation rate through Hurkx Trap-Assisted Tunneling (TAT) \cite{121690} and Poole-Frenkel effect (PF) \cite{COLALONGO1997627} is considered through the emission/capture cross section defined as

\begin{equation}
{\sigma _{t}=\sigma _{t}^{0} \cdot (1+\Gamma_{TAT,PF})}
\label{sigma}\end{equation}
 
where $\Gamma_{TAT,PF}$ is a field-enhancement factor defined with the default parameters specified in \cite{sdevice}. Taking into account equations \ref{G_a}-\ref{sigma} and utilizing the parameters provided in Table \ref{tab2}, the dark current (I$_{\text{d}}$) equals the product of G$_{\text{d}}$ and q, multiplied by the area occupied by defects V$_{\text{d}}$. Additionally, a second term representing the dark-induced multiplication current is incorporated, resulting in

\begin{equation}
I_d = q \cdot (G_d \cdot V_d + \alpha _{n,p} \cdot G_d \cdot \frac{V_d}{S_{SCR}})
\refstepcounter{equation}\tag{\theequation*} 
\label{IDEQ}\end{equation}

\begin{figure}[h]
\centerline{\includegraphics[width=\columnwidth]{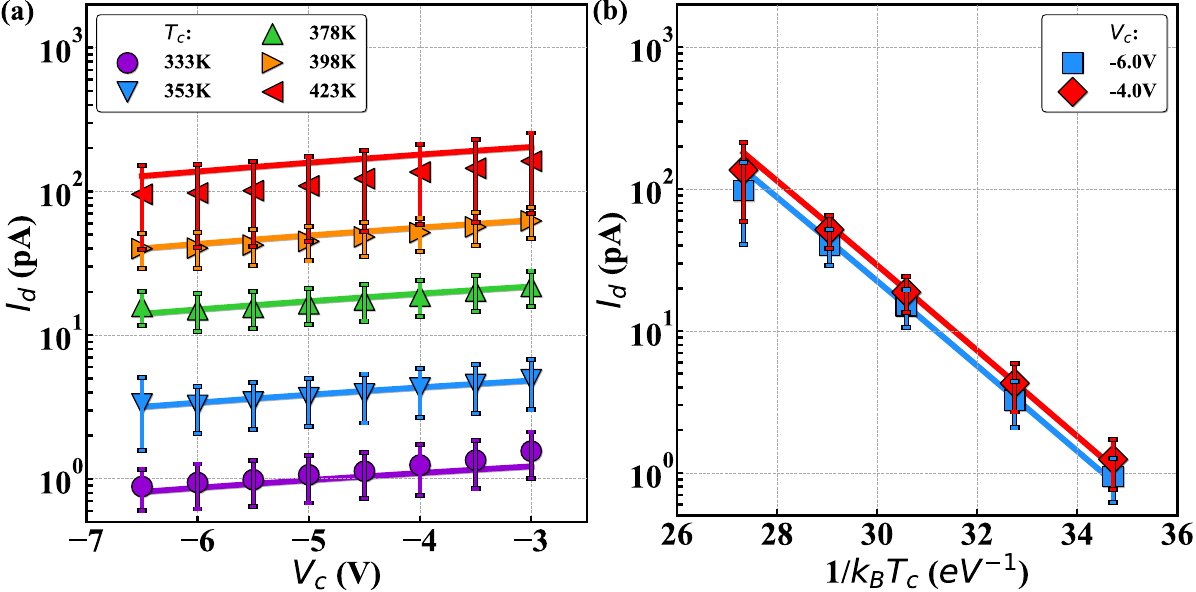}}
\caption{ (a) Dark current as a function of V$_{\text{c}}$ below V$_{\text{BD}}^{\text{LB}}$ measured (symbols) and modeled (lines) at different T$_{\text{c}}$. (b) Current as a function of 1/k$_{\text{B}}$T$_{\text{c}}$ measured (symbols) and modeled (lines) at different V$_{\text{c}}$. Error bars represent the standard deviation between each device.}
\label{IV_T_v1}
\end{figure}

\begin{figure}[h]
\centerline{\includegraphics[width=0.7\columnwidth]{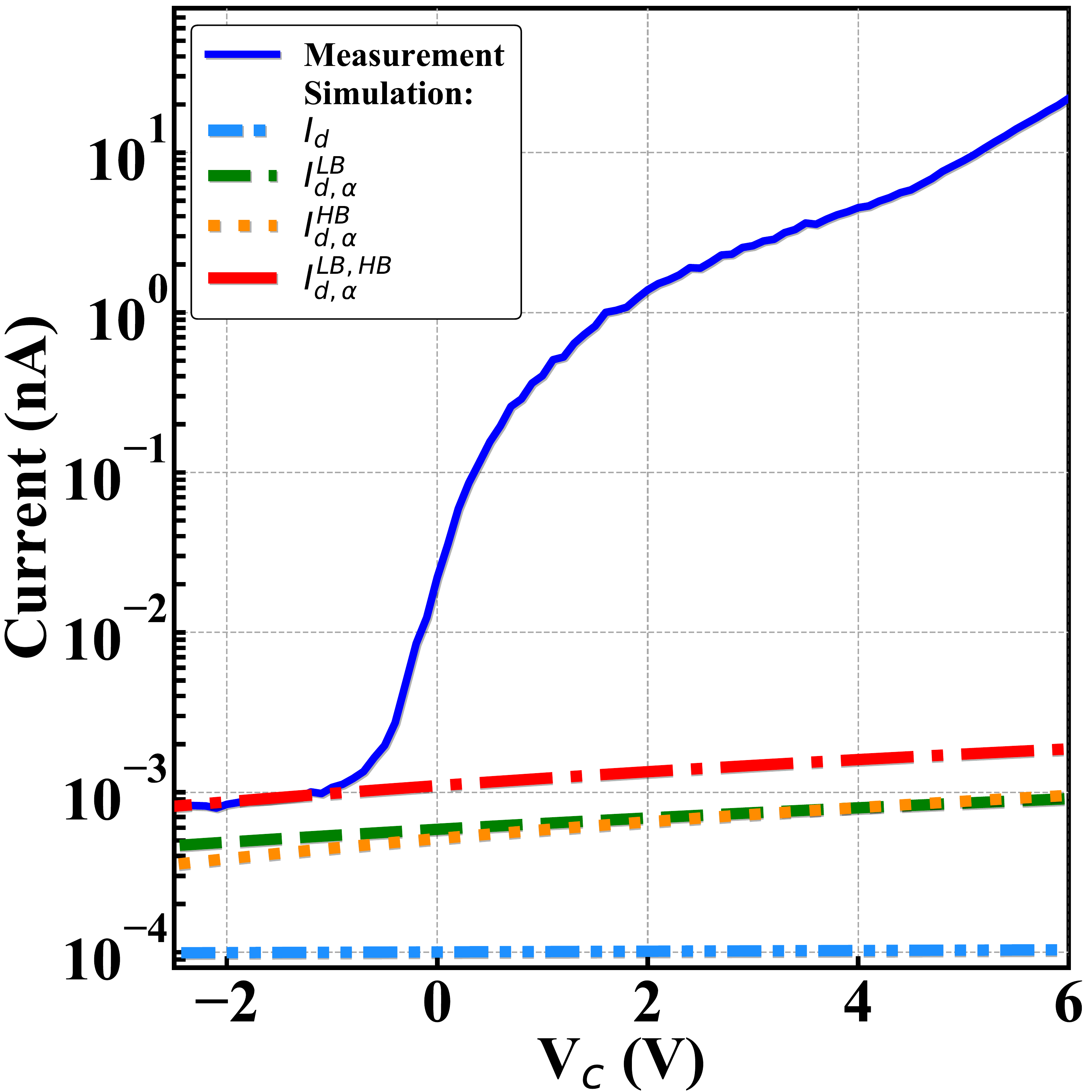}}
\caption{ Dark current as a function of V$_{\text{c}}$ below or above V$_{\text{BD}}^{\text{LB}}$ measured (bold lines) and modeled (dashed/dotted lines) considering dark current without or with carrier impact-ionization within LB, HB and both regions. Measuring current above V$_{\text{BD}}^{\text{LB}}$ without avalanche is impossible due to the strong avalanche effect. The discrepancy between the model and measurements above the breakdown voltage primarily stems from the presence of avalanche current.}
\label{IVT2}
\end{figure}

The dark current modeling reproduces experimental trend as shown along voltage in Fig. \ref{IV_T_v1}\textcolor{blue}{a} and temperature in Fig. \ref{IV_T_v1}\textcolor{blue}{b}. As previously mentioned, the strong avalanche effect makes it impossible to measure current above V$_{\text{BD}}^{\text{LB}}$ without avalanche. This is supported by dark current modeling results, as illustrated in Fig. \ref{IVT2}, where the current without avalanche is significantly lower than the current with avalanche. The dynamics of the voltage can be explained by the critical role played by multiplication caused by the impact-ionization of carriers.

\subsection{Photo-generated current modeling}

The photo-generation rate is modeled according to the Beer-Lambert law \cite{Swinehart} and assuming all photo-generated carriers are collected in the depletion region which leads to: 

\begin{equation}G_{ph}=q F_{0}(e^{-\alpha y_{0}}-e^{-\alpha y_{1}})\label{G_ph}\end{equation}

where $\alpha$ is the absorption coefficient per unit length, F$_{\text{0}}$ is the incident monochromatic photon flux (which is the irradiance reported in {{Table}} \ref{tab1} converted in number of photons per second and surface unit), and y$_{\text{0}}$ and  y$_{\text{1}}$ are the absorption depth at the surface and at the edge of the depleted volume respectively (Fig. \ref{SPAD_cut_DCR_T_v1}\textcolor{blue}{b}). The absorption coefficient $\alpha$ in indirect band gap semiconductor is proportional to \cite{Pankove_1972}:

\begin{equation}
\alpha \propto \frac{(h\nu - E_{g}+E_{p})^{2}}{exp(\frac{E_{p}}{k_{B}T})-1}+\frac{(h\nu - E_{g}-E_{p})^{2}}{1-exp(\frac{-E_{p}}{k_{B}T})}
\label{alpha}\end{equation}

The first term relates to phonon absorption and the second one to phonon emission with E$_{\text{p}}$ the phonon energy and $\nu$ the light frequency considering the light wavelength defined at 940nm. Taking into consideration equations \ref{G_ph}-\ref{alpha} and \ref{G_a}-\ref{gam}, along with the parameters provided in Table \ref{tab2}, the photo-generated current (I$_{\text{ph}}$) equals G$_{\text{ph}}$ multiplied by q and the effective area of the space charge region V$_{\text{SCR}}$. Additionally, a second term representing the photo-induced multiplication current is incorporated, resulting in:

\begin{equation}
I_{ph} = q \cdot (G_{ph} \cdot V_{SCR} + \alpha _{n,p} \cdot G_{ph} \cdot \frac{V_{SCR}}{S_{SCR}})
\refstepcounter{equation}\tag{\theequation*} 
\label{IPHEQ}\end{equation}

\begin{figure}[h]
\centerline{\includegraphics[width=\columnwidth]{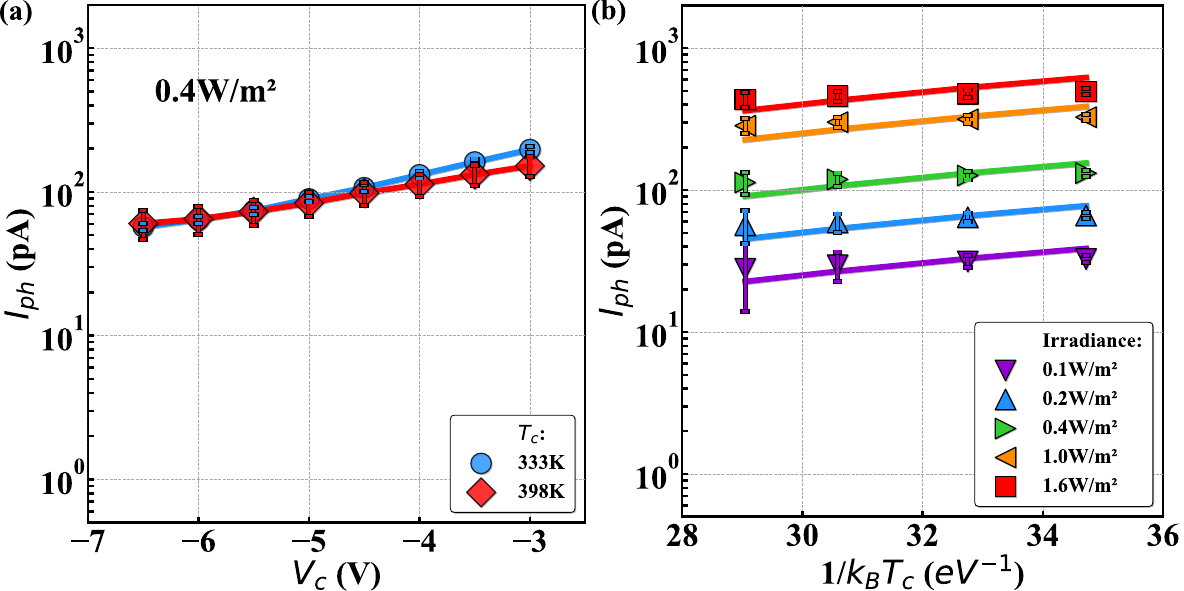}}
\caption{(a) Photo-generated current as a function of V$_{\text{s}}$ below V$_{\text{BD}}^{\text{LB}}$ (without avalanches) measured (symbols) and modeled (lines) at different T$_{\text{s}}$ and at 0.4W/m². (b) Current as a function of 1/k$_{\text{B}}$T$_{\text{s}}$ measured (symbols) and modeled (lines) at V$_{\text{s}}$=-4V for several irradiance levels. Error bars represent the standard deviation between each device. }
\label{IV_irr_T_v1}
\end{figure}

The photo-generated modeling shows good agreement with experiment as shown along voltage (Fig. \ref{IV_irr_T_v1}\textcolor{blue}{a}) and temperature (Fig. \ref{IV_irr_T_v1}\textcolor{blue}{b}) range. A distinction in current magnitude is observed above V$_{\text{c}}$=-5V and despite identical  irradiance level which relates to carrier multiplication in the depletion region as expected from electric field simulations (red line and orange box in Fig. \ref{E_HB_v1}\textcolor{blue}{a}).

A novel approach to electric current modeling was introduced, which is built upon existing works in the field: the SRH generation rate for dark current, the Beer-Lambert law for photo-generated current, and the Chynoweth law for carrier multiplication. Interactions between these different currents are considered in the modeling framework, with their dynamic interplay being taken into account under varying operational conditions. This integrated approach is utilized to provide a more accurate and comprehensive representation of the electric current dynamics in the context of SPAD devices. To the best of the authors' knowledge, this representation is offered for the first time, providing valuable insights that may not be readily obtainable through existing modeling methodologies. These approaches will be employed to quantify the population of hot carriers based on the carrier energy distribution and to model the degradation rate.

\section{Defect Creation Kinetics}

The defect creation dynamics must be explained to capture $\Delta$DCR along stress time accounting for the bond dissociation energy distributions modulated by hot-carrier collisions with bonds. The bond dissociation energy E$_{\text{hb}}$ depends on the local variation of electrical and chemical potential, such as those arising from dangling bond passivation by hydrogen. Disorder-induced variations among these Si-H bond dissociation energies E$_{\text{hb}}$ at the Si/SiO$_{\text{2}}$ interface have been shown \cite{doi:10.1063/1.116308,Stesmans} to be a plausible degradation source of the sub-linear time dependence of the defect creation process. Assuming that concentration of Si-H bond precursors N$_{\text{hb}}$(t) follows a first-order rate equation during stress and that defect creation events are independent, the corresponding portion of the interface defect density N$_\text{t}$(t) along stress time is found to be:

\begin{equation}
N_{t}(t)=N_{hb}^{0} (1-exp(-\kappa(t)t))
\refstepcounter{equation}\tag{\theequation*} 
\label{Ntt}\end{equation}

N$_{\text{hb}}^{0}$ is the initial concentration of Si-H bond precursors. The distribution of disorder-induced variations of Si-H bond-dissociation energies is assumed to follow a Fermi-derivative distribution \cite{doi:10.1063/1.357062,922913}:

\begin{equation}
\tilde{g}(E_{hb},t)=\frac{N_{hb}(t)}{\sigma_{hb}} \frac{exp(\frac{E_{th}-E_{hb}}{\sigma_{hb}})}{{[1+exp(\frac{E_{th}-E_{hb}}{\sigma_{hb}})]}^{2}} 
\refstepcounter{equation}\tag{\theequation*} 
\label{GG}\end{equation}

E$_{\text{th}}$ and $\sigma_{\text{hb}}$ are the mean and the Half Width at Half Maximum (HWHM) of the bond dissociation energy distribution respectively. The required bond-dissociation energy E$_{\text{hb}}$ to break a Si-H bond does not correspond to the defect energy level E$_{\text{t}}$ resulting from this process. The change in $\tilde{\text{g}}(\text{E}_{\text{hb}}\text{,t})$ along stress time is considered by the concentration of Si-H bond precursors N$_{\text{hb}}$(t) that reduces proportionally with the creation of dangling bond defects N$_{\text{t}}$(t), following the relationship N$_{\text{hb}}$(t) = N$_{\text{hb}}^0$ - N$_{\text{t}}$(t). The bond dissociation rate constant $\kappa$ must account for the cumulative ability of the carrier ensemble to dissociate bonds by impact-ionization and for the Arrhenius process expressing that reactions proceed exponentially faster as the temperature is raised which results in:

\begin{equation}
\centering
\begin{split}
\kappa(t)= & \int_{0}^{+\infty} \int_{E_{hb}}^{+\infty} \frac{1}{q} \cdot (j_{d,\alpha}^{total}(t) + j_{ph,\alpha}^{total}) \cdot \frac{n(E)}{n_{total}} \\
& \cdot  \sigma_{II}(E,E_{hb}) \frac{\tilde{g}(E_{hb},t)}{\tilde{g}_{total}(t)} exp(\frac{-E_{hb}}{k_{b}T}) dE dE_{hb}
\end{split}
\refstepcounter{equation}\tag{\theequation*} 
\label{KK}\end{equation}

where the integration is performed over the bond energy E$_{\text{hb}}$ and over the carrier energy E starting from E$_{\text{hb}}$. n(E) is integrated over the degradation area. n$_{\text{total}}$ stands for the total carrier energy distribution and $\tilde{\text{g}}_{\text{total}}(\text{t})$ for the total Si-H bond dissociation energy distribution along stress duration. The Keldysh-like impact-ionization reaction cross section $\sigma_{II}=\sigma_{II}^{0}(E-E_{hb})^{\rho_{0}}$ is used for the SP mechanism triggered by hot carriers with $\sigma_{\text{II}}$ being the attempt rate and $\rho_{0}$ the energy exponent of the impact-ionization reaction dynamics fitted on HCD in NMOS \cite{doi:10.1063/1.351434}. The attempt rate $\sigma_{\text{II}}^{0}$ drives the degradation temporal dynamics and the number of defect precursors N$_{\text{hb}}$ the degradation magnitude. This model gives a comprehensive explanation of the sub-linear time-dependence of the degradation driven by the Si-H bond dissociation energy distribution downsizing by the hot-carriers impact-ionizing the bonds. The model predicts the saturation of defect creation when time tends to infinity due to the exhaustion of defect precursors N$_{\text{hb}}$ \cite{Blat}.

\begin{figure}[h]
\centerline{\includegraphics[width=\columnwidth]{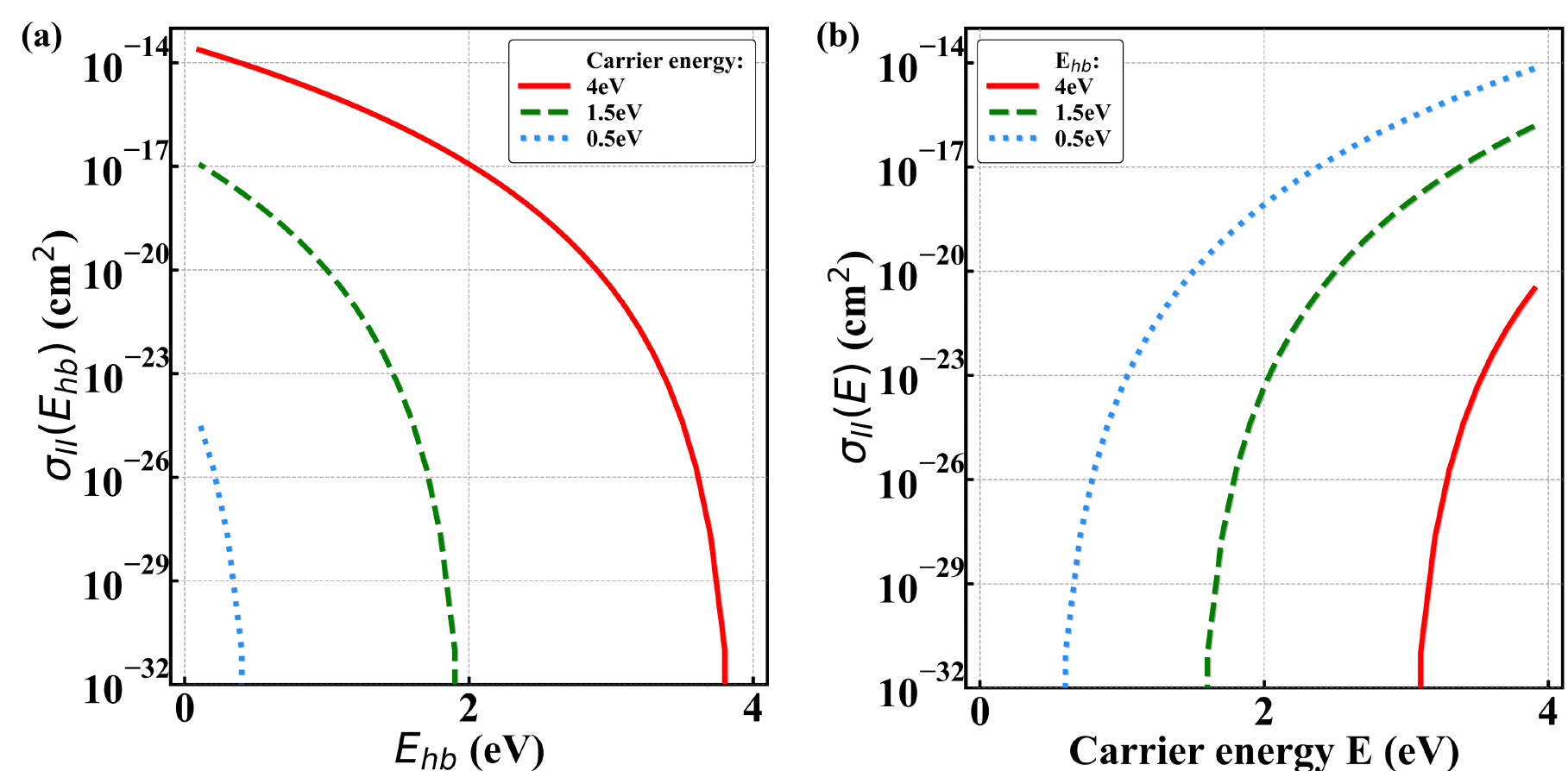}}
\caption{$\sigma_{\text{II}}$ as a function of (a) the Si-H bond energy E$_{\text{hb}}$ and (b) the carrier energy E from low to high E and E$_{\text{hb}}$ respectively. The impact-ionization probability of dissociating a Si-H bond decreases with increasing Si-H bond energy but increases with increasing carrier energy. The increase in the creation of defects at a given stress time corresponds to the bond-dissociation energy states that are available for impact-ionization by hot carriers.}
\label{sigEEhb}
\end{figure}

\begin{figure}[h]
\centerline{\includegraphics[width=\columnwidth]{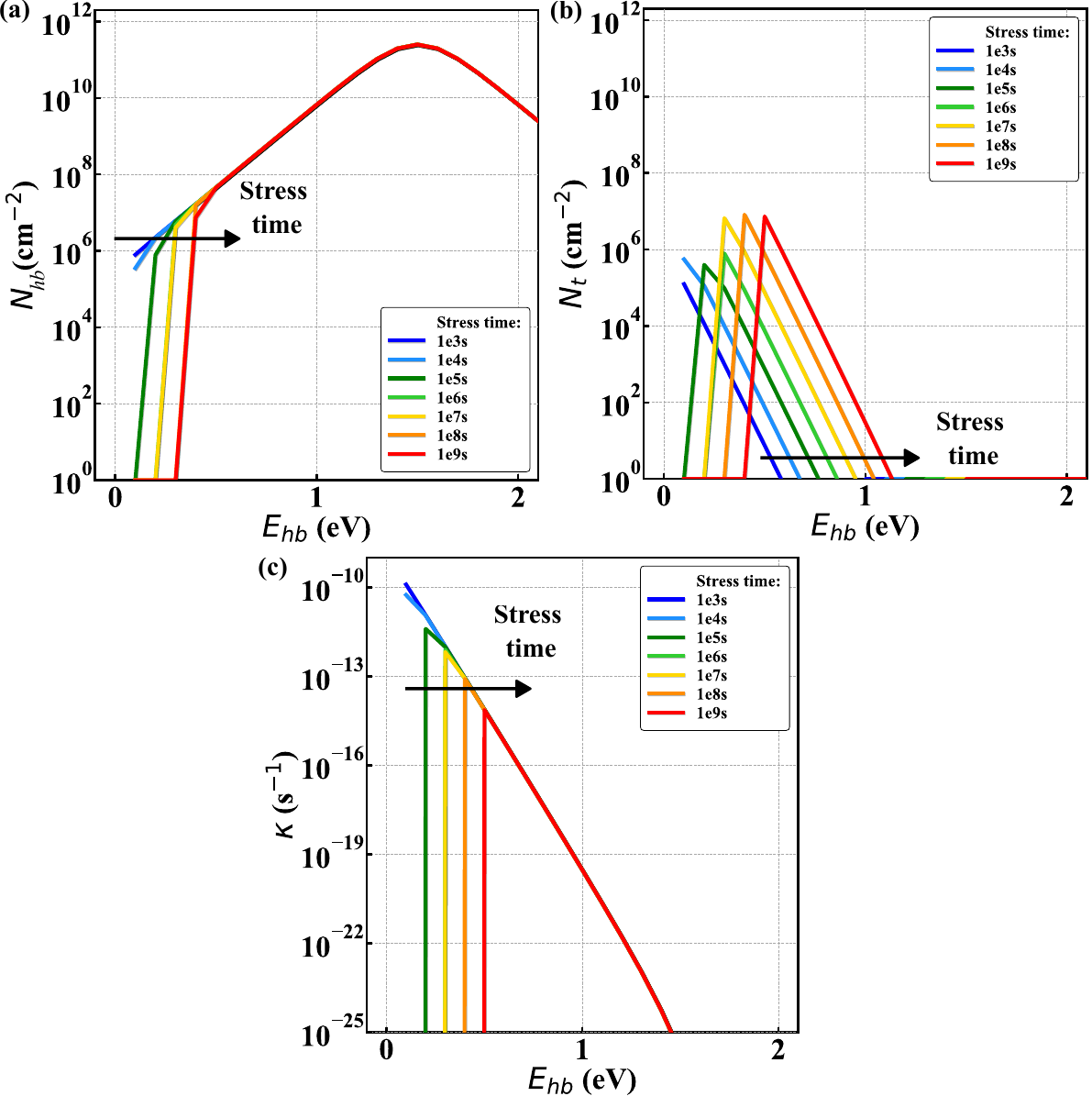}}
\caption{(a) Si-H bond-dissociation energy distribution N$_{\text{hb}}$,(b) defect distribution N$_{\text{t}}$ and (c) bond dissociation rate constant $\kappa$ as a function of the Si-H bond energy E$_{\text{hb}}$ along stress duration. The impact-ionization probability results in the depletion of Si-H bond-dissociation energy states at low E$_{\text{hb}}$.}
\label{sigEEhb2}
\end{figure}

In Fig. \ref{sigEEhb}, $\sigma_{\text{II}}$ is plotted as a function of both the Si-H bond energy E$_{\text{hb}}$ and the carrier energy E, with the data ordered from low to high values of each parameter. As the Si-H bond energy increases, the probability of dissociating a Si-H bond by impact-ionization decreases, while it increases with increasing carrier energy. As a result, the distribution of Si-H bond-dissociation energy states becomes exhausted at low E$_{\text{hb}}$, as shown in Fig. \ref{sigEEhb2}\textcolor{blue}{a}. Additionally, there is an increase in the creation of defects at a given stress time, which corresponds to the bond-dissociation energy states that are available for impact-ionization by hot carriers, as depicted in Fig. \ref{sigEEhb2}\textcolor{blue}{b}. Consequently, the bond dissociation rate constant $\kappa$ undergoes a change over the duration of the stress, as depicted in Fig. \ref{sigEEhb2}\textcolor{blue}{c}.

\section{Dark Count Rate Drift modeling}

After identifying the defect formation kinetics, the generation rate from the stress-induced defects must be calculated. The same previous hypothesis stands for the defect creation  \cite{SICRE3} regarding the initial defect generation rate of carriers ({{eq.}} \ref{Gd}) at the top depleted interface and at a single-energy mide-gap level. At the considered defect density N$_{\text{t}}$, the presence of traps inside the device does not significantly alters the device electrostatic profile \cite{SICRE1} thus P$_{\text{t}}$ is assumed to be constant along stress time. According to these assumptions, the DCR created along stress time can be rewritten in a simple form by replacing N$_{\text{t}}^{\text{0}}$ by N$_{\text{t}}$(t) in G$_{\text{d}}$(t) ({{eq.}} \ref{Gd}) as follows:

\begin{equation}
\begin{split} 
DCR(t)= & P_{t} \cdot G_{d}(t) = P_{t} \cdot N_{hb}^{0} (1-exp(-\kappa(t)t)) \\
& \cdot \frac{n_{i} \sigma_{t} v_{th}^{n} v_{th}^{p}}{v_{th}^{n} exp(\frac{E_{t}-E_{F}^{i}}{k_{B}T}) + v_{th}^{p} exp(\frac{E_{F}^{i}-E_{t}}{k_{B}T})}
\end{split}
\refstepcounter{equation}\tag{\theequation*} 
\label{DCR_t}\end{equation}

\begin{figure}[h]
\centerline{\includegraphics[width=\columnwidth]{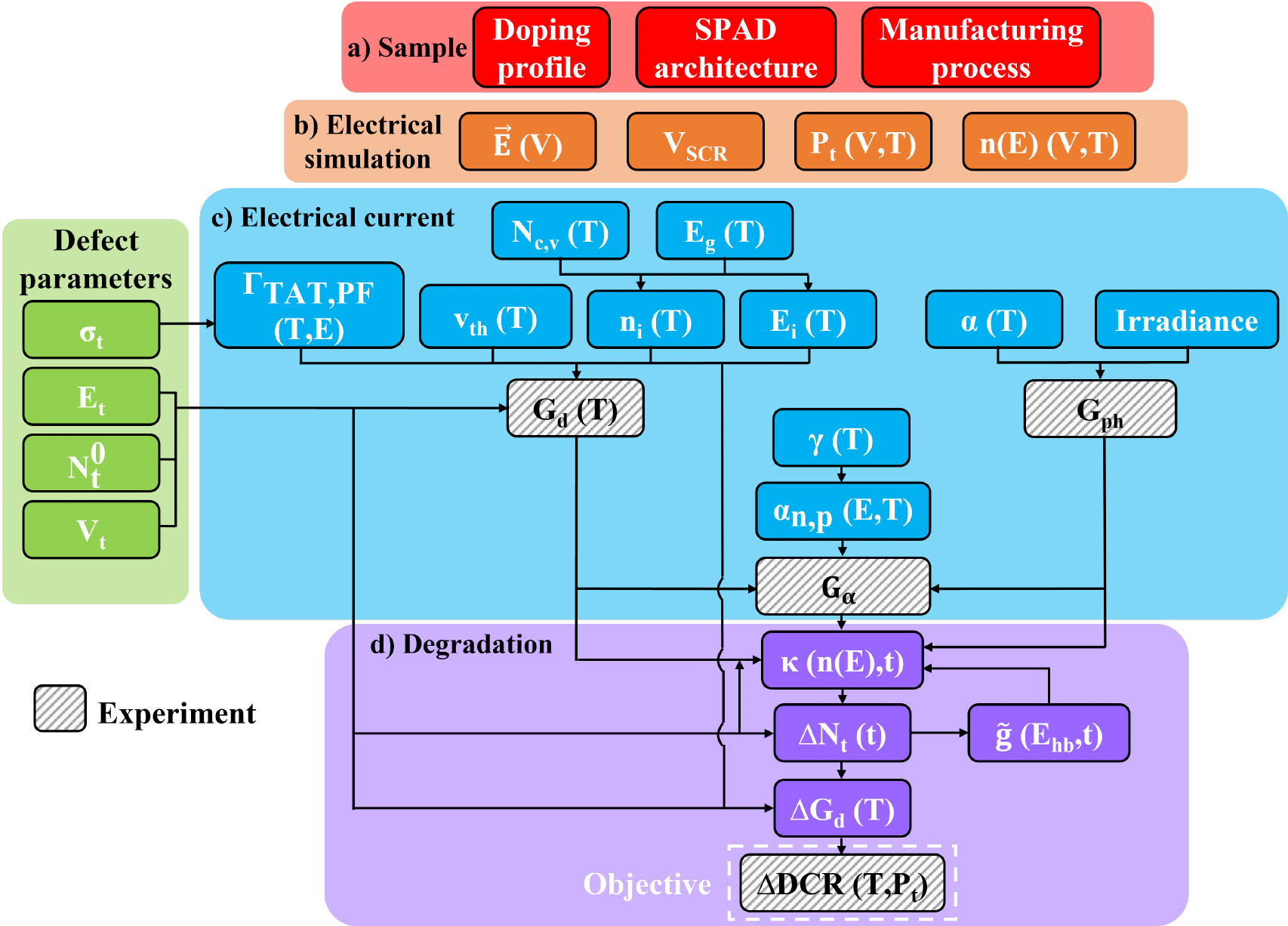}}
\caption{Proposed methodology to predict $\Delta$DCR based on: (a) SPAD architecture and process to simulate (b) electric field $\overrightarrow{\text{E}}$, volume of space charge region V$_{\text{SCR}}$, avalanche breakdown probability P$_{\text{t}}$ and carrier energy distribution n(E) integrated with (c) dark-, photo- and multiplication rates G$_{\text{d},\text{ph},\alpha}$ to calculate (d) the bond dissociation rate constant $\kappa$. See parameter description in the text. Only first order dependencies are shown in parenthesis such as the bond dissociation rate constant $\kappa$ which depends on the energy carrier distribution n$_{\text{E}}$ (first order) which depends on V and T (second order). Defect parameters are arbitrarily fixed. Variables in dashed grey boxes are fitted with experimental data.}
\label{Methodology_v1}
\end{figure}

The modeling methodology is summarized in Fig. \ref{Methodology_v1}. The characterization effect is interpreted from the first and the third term of {{eq.}} \ref{DCR_t} which are the avalanche breakdown probability P$_{\text{t}}$ and the carrier generation rate from created defects along stress duration divided by the number of created defects. P$_{\text{t}}$ is simulated along V$_{\text{c}}$ and T$_{\text{c}}$ (Fig. \ref{Pt_TV_v1}). The intrinsic concentration n$_{\text{i}}$ is calculated from {{eqs.}} \ref{n_i}-\ref{E_i} and the thermal velocity v$_{\text{th}}$ from {{eqs.}} \ref{v_th} as a function of T$_{\text{c}}$. The emission/capture cross section $\sigma _{\text{t}}$ is computed from {{eq.}} \ref{sigma} as a function of V$_{\text{c}}$ depending on the simulated electric field $\overrightarrow{\text{E}}$ (Fig. \ref{E_HB_v1}\textcolor{blue}{a}). n$_{\text{i}}$, v$_{\text{th}}$ and $\sigma _{\text{t}}$ are injected into {{eqs.}} \ref{Gd} to calculate the generation rate of carriers from stress-induced defects.

The stress impact is captured by the second term of {{eq.}} \ref{DCR_t} which is the number of defects N$_{\text{t}}$(t) created along stress time from {{eq.}} \ref{Ntt}. The time-dependent bond dissociation rate constant $\kappa$(t) is computed from {{eq.}} \ref{KK}. The carrier energy distribution n(E) is simulated by the Full-Band Monte-Carlo ({{Section}} \ref{sec:EC}) along V$_{\text{s}}$ and  T$_{\text{s}}$ range. The dark- and photo-generated current densities are modeled from {{eqs.}} \ref{Gd} (Fig. \ref{IV_T_v1}) and \ref{G_ph} (Fig. \ref{IV_irr_T_v1}) respectively coupled with the multiplication current from {{eq.}} \ref{G_a} considering the simulated electric field $\overrightarrow{\text{E}}$ (Fig. \ref{E_HB_v1}\textcolor{blue}{a}). The Si-H bond dissociation energy distribution $\tilde{\text{g}}(\text{E}_{\text{hb}}\text{,t})$ is calculated from {{eq.}} \ref{GG} at each iteration to account for the exhaustion of defect precursors by dangling-bond defects.

\begin{figure}[h]
\centerline{\includegraphics[width=\columnwidth]{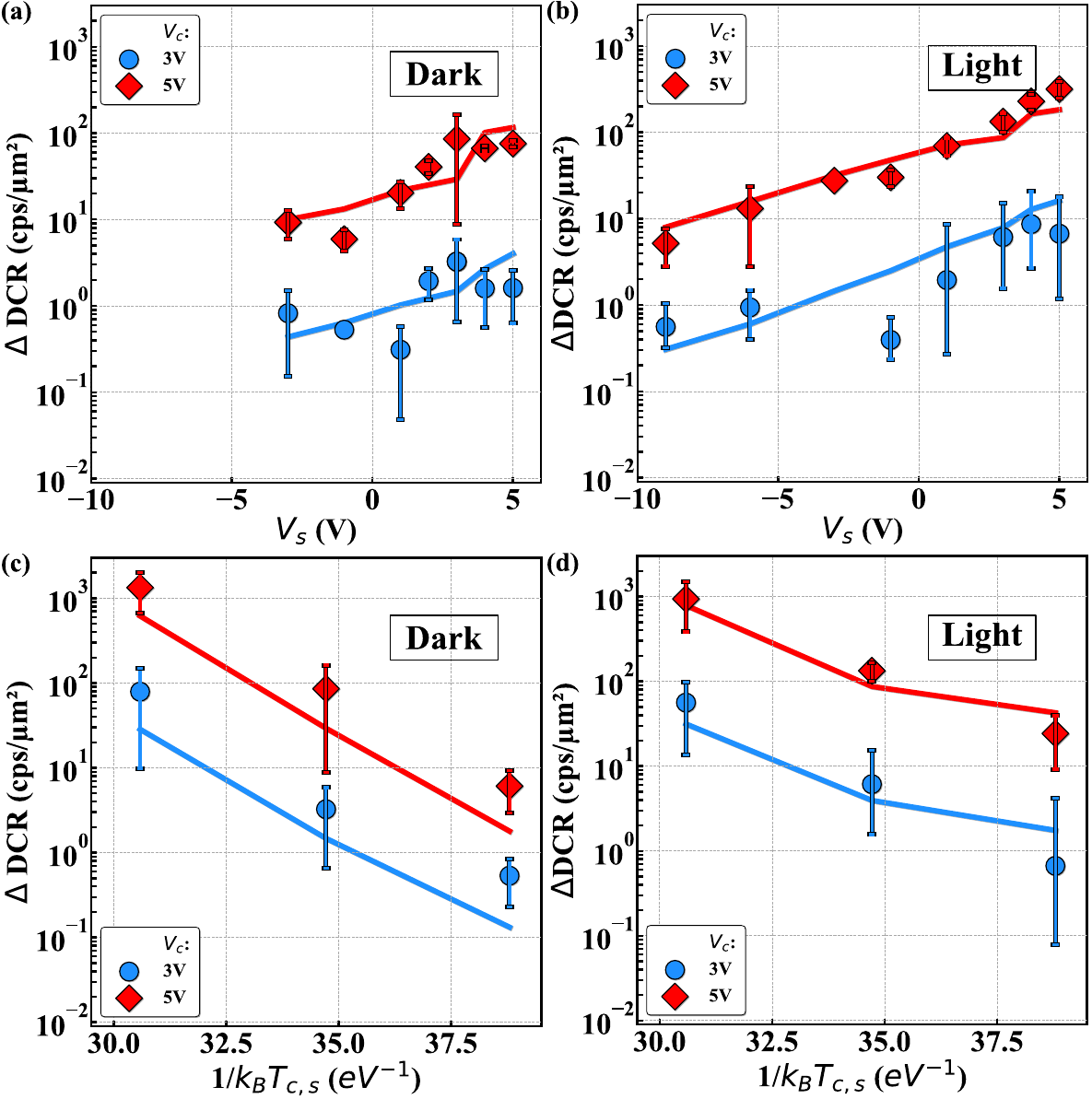}}
\caption{$\Delta$DCR as a function of V$_{\text{s}}$ for stresses (a) in darkness and (b) at 0.3W/m² for T$_{\text{s}}$=333K measured (symbols) and modeled (lines) at T$_{\text{c}}$=333K and at V$_{\text{c}}$=3V (LB - blue) and 5V (HB - blue). $\Delta$DCR as a function of 1/k$_{\text{B}}$T$_{\text{s}}$ for stresses (c) in darkness and (d) at 0.3W/m² for V$_{\text{s}}$=3V measured (symbols) and modeled (lines) at T$_{\text{c}}$=T$_{\text{s}}$ and at V$_{\text{c}}$=3V (LB - blue) and 5V (HB - red). Error bars represent the standard deviation between each device for the same stress condition. The displayed experimental data points are extracted from \cite{SICREESS}.}
\label{DDCR_d_t_v1}
\end{figure}

This methodology (Fig. \ref{Methodology_v1}) is now confronted to experimental results. The degradation model is firstly compared to various stress voltage and temperature conditions across different characterization voltage and temperature conditions. For stresses in darkness and under illumination at different stress voltages and a constant stress temperature (first and second rows in {{Table}} \ref{tab1}), a good correlation is obtained between the proposed model and the experiment as shown in Fig. \ref{DDCR_d_t_v1}\textcolor{blue}{a-b}. From characterization perspective, the voltage dependence of avalanche breakdown probability enables the replication of the characterization trend in both the LB (V$_{\text{c}}$=3V) and HB (V$_{\text{c}}$=5V) regions. From stress perspective, the voltage dependence of hot-carrier multiplication enables the replication of the experimental trend in both dark and irradiance conditions. 

For stresses in darkness and under illumination at different stress temperatures and a constant stress voltage, a good correlation is obtained between the proposed model and the experiment as shown in Fig. \ref{DDCR_d_t_v1}\textcolor{blue}{c-d}. The characterization trend in the LB (V$_{\text{c}}$=3V) and HB (V$_{\text{c}}$=5V) regions is captured by the exponential thermal dependence of SRH generation rate. Disregarding temperature characterization effect due to the SRH generation rate of carriers which doubles each 10K in the temperature difference and the avalanche breakdown probability P$_{\text{t}}$ each 40K, the number of defects N$_{\text{t}}$(t) created along stress time doubles each 40K and 30K in darkness and under light respectively. The degradation increases as the temperature increases which is contrary to the behavior expected for generic SP in long-channel MOSFETs \cite{HONG1999809}. The degradation rates depend on the carrier energy and density. The carrier energy loss dynamics by scattering mechanisms increasing with temperature are the same from MOSFETs to SPADs as shown in Fig. \ref{Carrier_distrib_v1}\textcolor{blue}{d-e}. However in the MOSFETs transistor, the carrier density is proportional to the gate voltage minus the threshold voltage (which is approximately linear with temperature) whereas in SPADs is following the diode reverse current (which increases exponential with temperature through the SRH generation rate) as shown in Fig. \ref{IV_T_v1}\textcolor{blue}{b}.

\begin{figure}[h]
\centerline{\includegraphics[width=\columnwidth]{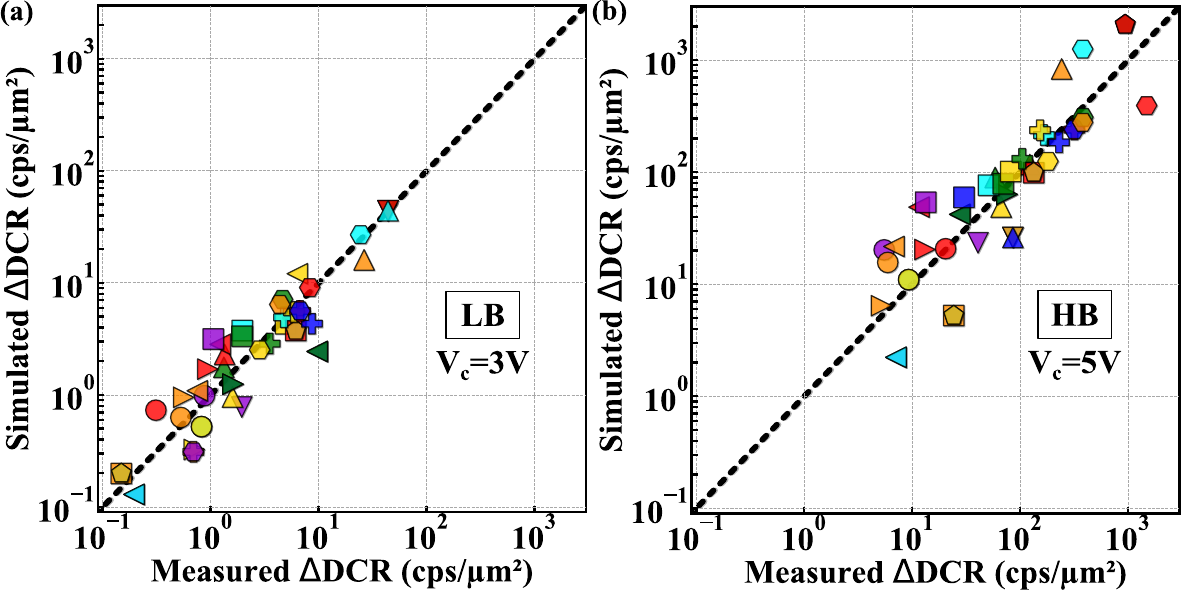}}
\caption{ $\Delta$DCR correlation plots between experiment and modeling measured at (a) V$_{\text{c}}$=3V (LB) and (b) 5V (HB) for each stress condition reported in {{Table}} \ref{tab1}. Each symbol corresponds to a different stress conditions averaged on eight devices. The dashed line is the y=x line to underline the correlation between experiment and modeling.}
\label{XY_v1}
\end{figure}

Furthermore, the correlation plots of LB and HB region shown in Fig. \ref{XY_v1} enlighten the appropriate relationship achieved for all stress conditions reported in {{Table}} \ref{tab1}.

\begin{figure}[h]
\centerline{\includegraphics[width=0.5\columnwidth]{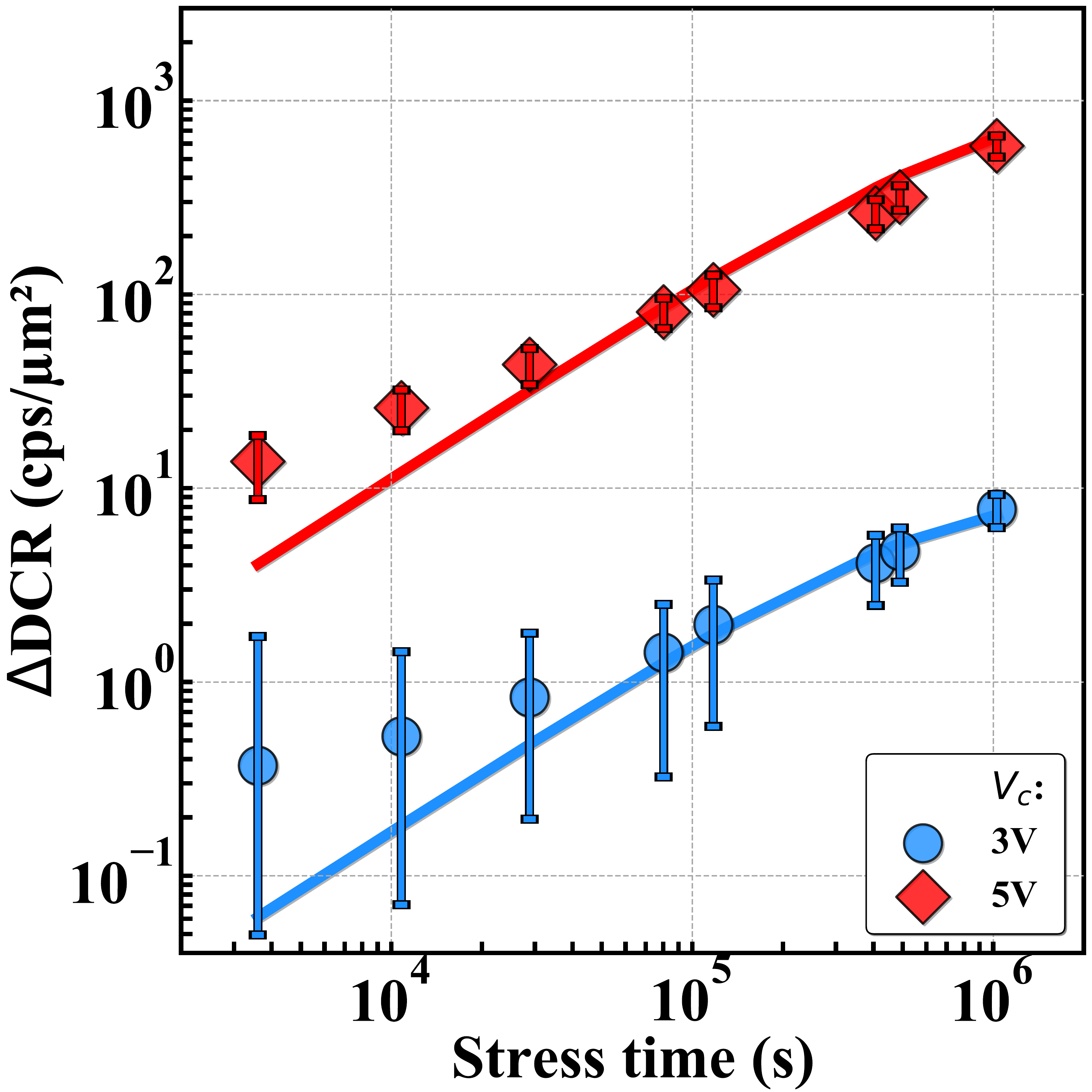}}
\caption{$\Delta$DCR as a function of the stress time for stresses on 64 SPADs in darkness at V$_{\text{s}}$=3V, V$_{\text{s}}$=353K and at 0.3W/m² measured (symbols) and modeled (lines) at V$_{\text{c}}$=3V (LB - blue) and 5V (HB - red) for T$_{\text{c}}$=333K.}
\label{DDCR_t_v1}
\end{figure}

In the existing literature, HCD have been extensively studied under short stress time to asymptotically approximate the solution of the first order kinetic equation by a power law \cite{Haggag} which is to said that created defect do not modify significantly the bond dissociation rate constant $\kappa$. In this study, the stress time range is extended up to 10$^{6}$s to probe long-term device aging behavior. The modeled stress temporal dynamics characterized in both region is in good agreement with the experiment as shown in Fig. \ref{DDCR_t_v1}. The time-dependent degradation is effectively captured by considering the time dependence of the first-order kinetics equation for Si-H bond dissociation, which incorporates the modulation of the bond dissociation rate constant by the exhaustion of Si-H defect precursors. Nevertheless, the slight discrepancy noticed at low stress times below $10^4$s, where it may appear that there are two trends, may arise from the use of a single Si-H bond distribution to model degradation kinetics, which may not be accurate for the micrometer squared area under investigation. Introducing multiple distributions that are stochastically addressed could potentially enhance the accuracy of the model.

Additionally, measuring low count rate drift can be difficult due to the statistical dispersion from device-to-device, where the $\Delta$DCR mean follows a quadratic law with its variance \cite{SICRE4}. 

\section{Dark current drift modeling}

The determination of the generation rate from stress-induced defects can also aid in assessing the drift in dark current over the duration of the stress. This dark current along stress time is established by substituting N$_{\text{t}}^{\text{0}}$ with N$_{\text{t}}$(t) in equation eq. \ref{Gd}, while keeping all other parameters constant, together with eq. \ref{IDEQ}. The hypothesis previously stated regarding defect creation also applies to the initial generation rate of carriers (eq. \ref{Gd}) at the top depleted interface and at a single-energy mid-gap level E$_{\text{t}}$. These assumptions yield to:

\begin{equation}
\begin{split} 
I_d (t) = & q \cdot N_{hb}^{0} (1-exp(-\kappa(t)t)) \\
& \cdot \frac{n_{i} \sigma_{t} v_{th}^{n} v_{th}^{p}}{v_{th}^{n} exp(\frac{E_{t}-E_{F}^{i}}{k_{B}T}) + v_{th}^{p} exp(\frac{E_{F}^{i}-E_{t}}{k_{B}T})}  ( V_d + \alpha _{n,p} \cdot \frac{V_d}{S_{SCR}})
\end{split} 
\refstepcounter{equation}\tag{\theequation*} 
\label{IDEQ}\end{equation}

It is observed that the dark current increased proportionally with the number of defects created over time, as demonstrated in Fig. \ref{IV_T_v1}. This implies that the defect profile and localization remain unchanged before and after stress, similar to $\Delta$DCR. These findings suggest that the inclusion of dark current drift in the calculation of $\kappa$ using equation eq. \ref{KK} is necessary.

\begin{figure}[h]
\centerline{\includegraphics[width=\columnwidth]{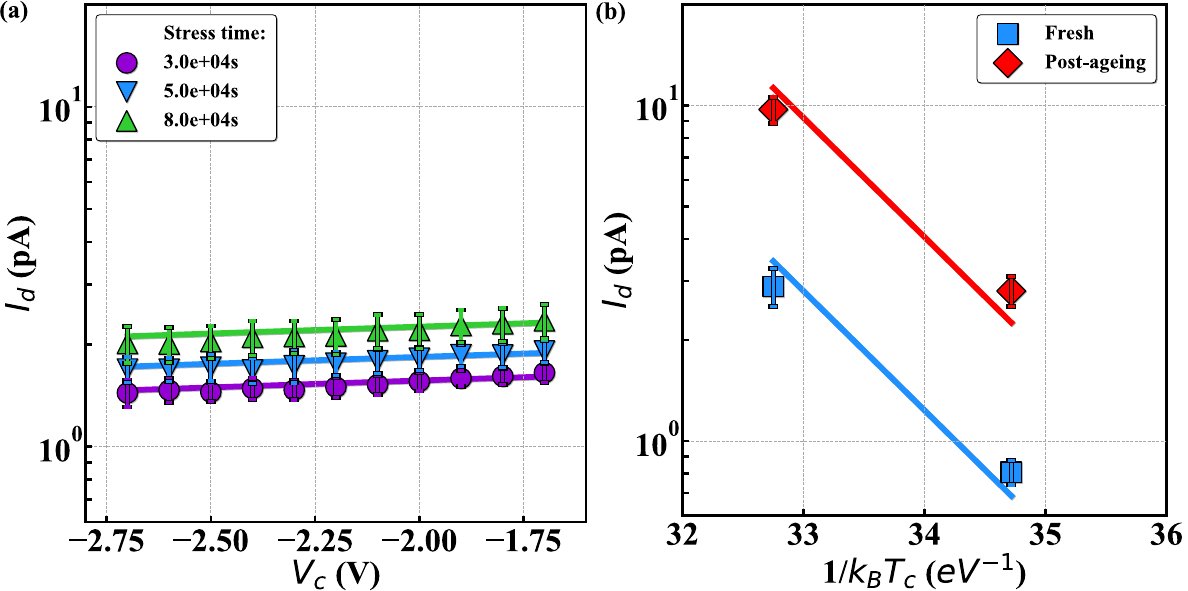}}
\caption{ (a) Dark current as a function of V$_{\text{c}}$ below V$_{\text{BD}}^{\text{LB}}$ measured (symbols) and modeled (lines) at different T$_{\text{c}}$ for an increasing stress duration. (b) Dark current as a function of 1/k$_{\text{B}}$T$_{\text{c}}$ measured (symbols) and modeled (lines) at V$_{\text{c}}$=-2V before (rectangles) and after (diamonds) stress. The stress was performed at stress voltage V$_{\text{s}}$=5V, irradiance of 0.3W/m² at 940nm and temperature T$_{\text{s}}$=373K. Error bars represent the standard deviation between each device. The assumption is made that the thermal dynamics are identical to the ones depicted in Figure \ref{IV_T_v1}\textcolor{blue}{b}.}
\label{ID_pre_post_v1}
\end{figure}

\section{Conclusion}
The developed model is capable of accurately describing the hot-carrier degradation that induces $\Delta$DCR in SPADs across a wide range of operating conditions. The number of available hot carriers is estimated by analyzing the current density in both darkness and under irradiance, using a full-band Monte-Carlo simulation to determine the carrier energy distribution. The main contributors to the total defect creation are identified as hot electrons. The sub-linear time dependence of the defect creation process is captured by the exhaustion of the bond dissociation energy distribution due to impact-ionizing hot carriers. The two different slopes observed in hot-carrier degradation are explained by the degradation occurring in two different regions of the device architecture. In contrast to long-channel MOSFETs, the carrier density in SPADs increases exponentially with the applied voltage, leading to an increase in degradation as a function of temperature. The model presented is capable of accurately predicting the increase in dark current that occurs during stress. This is achieved by incorporating the dark current drift into the calculation of $\kappa$. The methodology can be used to predict device performance lifetime depending on characterization and stress conditions to ensure image integrity.


\small
\section*{Copyright notice}
Figures \ref{SPAD_cut_DCR_T_v1}, \ref{E_HB_v1}\textcolor{blue}{a}, \ref{Pt_TV_v1}\textcolor{blue}{a}, \ref{Carrier_distrib_v1}\textcolor{blue}{a-b}, \ref{IV_T_v1}\textcolor{blue}{a} and \ref{IV_irr_T_v1}\textcolor{blue}{a} are reprinted, with permission, from M. Sicre, D. Roy, and F. Calmon, “Hot-Carrier Degradation modeling of DCR drift in SPADs,” in ESSDERC 2023 - IEEE 53rd European Solid-State Device Research Conference (ESSDERC) (2023) pp. 61–64 \cite{SICREESS}. © 2023 IEEE.

\small
\bibliography{SPAD_DCR_HCD_Sicre_Federspiel_Mamdy_Roy_Calmon}

\begin{thebibliography}{73}
\providecommand{\natexlab}[1]{#1}
\providecommand{\url}[1]{#1}
\csname url@samestyle\endcsname
\providecommand{\newblock}{\relax}
\providecommand{\bibinfo}[2]{#2}
\providecommand{\BIBentrySTDinterwordspacing}{\spaceskip=0pt\relax}
\providecommand{\BIBentryALTinterwordstretchfactor}{4}
\providecommand{\BIBentryALTinterwordspacing}{\spaceskip=\fontdimen2\font plus
\BIBentryALTinterwordstretchfactor\fontdimen3\font minus \fontdimen4\font\relax}
\providecommand{\BIBforeignlanguage}[2]{{%
\expandafter\ifx\csname l@#1\endcsname\relax
\typeout{** WARNING: IEEEtranN.bst: No hyphenation pattern has been}%
\typeout{** loaded for the language `#1'. Using the pattern for}%
\typeout{** the default language instead.}%
\else
\language=\csname l@#1\endcsname
\fi
#2}}
\providecommand{\BIBdecl}{\relax}
\BIBdecl

\bibitem[Piron et~al.(2021)Piron, Morrison, Yuce, and Redouté]{9264255}
F.~Piron, D.~Morrison, M.~R. Yuce, and J.-M. Redouté, ``{A Review of Single-Photon Avalanche Diode Time-of-Flight Imaging Sensor Arrays},'' \emph{IEEE Sensors Journal}, vol.~21, no.~11, pp. 12\,654--12\,666, 2021.

\bibitem[Dumas et~al.(2024)Dumas, Coughlan, Smith, Mirza, Kirdoda, Fleming, McCarthy, Mowbray, Yi, Saalbach, Buller, Paul, and Millar]{10.1117/12.3001450}
D.~C.~S. Dumas, C.~Coughlan, C.~Smith, M.~Mirza, J.~Kirdoda, F.~Fleming, C.~McCarthy, H.~Mowbray, X.~Yi, L.~Saalbach, G.~S. Buller, D.~J. Paul, and R.~W. Millar, ``{Decoupling the dark count rate contributions in Ge-on-Si single photon avalanche diodes},'' in \emph{Quantum Sensing and Nano Electronics and Photonics XX}, M.~Razeghi, G.~A. Khodaparast, and M.~S. Vitiello, Eds., vol. 12895, International Society for Optics and Photonics.\hskip 1em plus 0.5em minus 0.4em\relax SPIE, 2024, p. 1289505.

\bibitem[Qian et~al.(2023)Qian, Jiang, Elsharabasy, and Deen]{s23073412}
X.~Qian, W.~Jiang, A.~Elsharabasy, and M.~J. Deen, ``Modeling for single-photon avalanche diodes: State-of-the-art and research challenges,'' \emph{Sensors}, vol.~23, no.~7, 2023.

\bibitem[Liu et~al.(2023)Liu, Li, Xu, Dong, Fang, and Xu]{Liu_2023}
D.~Liu, M.~Li, T.~Xu, J.~Dong, Y.~Fang, and Y.~Xu, ``Study of the influence of virtual guard ring width on the performance of spad detectors in 180 nm standard cmos technology,'' \emph{Journal of Semiconductors}, vol.~44, no.~11, p. 114102, nov 2023.

\bibitem[Bian et~al.(2023)Bian, Liu, Dong, Yuan, and Xu]{Dajing}
D.~Bian, D.~Liu, J.~Dong, F.~Yuan, and Y.~Xu, ``A p-type enrichment/high voltage n-well junction si-spad with enhanced near-infrared sensitivity in 180 nm bcd technology,'' \emph{IEEE Electron Device Letters}, vol.~PP, pp. 1--1, 01 2023.

\bibitem[Panglosse et~al.(2020)Panglosse, Martin-Gonthier, Marcelot, Virmontois, Saint-Pé, and Magnan]{8999742}
A.~Panglosse, P.~Martin-Gonthier, O.~Marcelot, C.~Virmontois, O.~Saint-Pé, and P.~Magnan, ``Dark count rate modeling in single-photon avalanche diodes,'' \emph{IEEE Transactions on Circuits and Systems I: Regular Papers}, vol.~67, no.~5, pp. 1507--1515, 2020.

\bibitem[Xu et~al.(2016)Xu, Xiang, Xie, and Huang]{Xu_2016}
Y.~Xu, P.~Xiang, X.~Xie, and Y.~Huang, ``A new modeling and simulation method for important statistical performance prediction of single photon avalanche diode detectors,'' \emph{Semiconductor Science and Technology}, vol.~31, no.~6, p. 065024, may 2016.

\bibitem[Sicre et~al.(2021)Sicre, Agnew, Buj, Coignus, Golanski, Helleboid, Mamdy, Nicholson, Pellegrini, Rideau, Roy, and Calmon]{SICRE1}
M.~Sicre, M.~Agnew, C.~Buj, J.~Coignus, D.~Golanski, R.~Helleboid, B.~Mamdy, I.~Nicholson, S.~Pellegrini, D.~Rideau, D.~Roy, and F.~Calmon, ``{Dark Count Rate in Single-Photon Avalanche Diodes: Characterization and Modeling study},'' in \emph{ESSCIRC 2021 - IEEE 47th European Solid State Circuits Conference (ESSCIRC)}, 2021, pp. 143--146.

\bibitem[Sicre et~al.(2022{\natexlab{a}})Sicre, Agnew, Buj, Coutier, Golanski, Helleboid, Mamdy, Nicholson, Pellegrini, Rideau, Roy, and Calmon]{SICRE2}
M.~Sicre, M.~Agnew, C.~Buj, C.~Coutier, D.~Golanski, R.~Helleboid, B.~Mamdy, I.~Nicholson, S.~Pellegrini, D.~Rideau, D.~Roy, and F.~Calmon, ``{Statistical measurements and Monte-Carlo simulations of DCR in SPADs},'' in \emph{ESSCIRC 2022- IEEE 48th European Solid State Circuits Conference (ESSCIRC)}, 2022, pp. 193--196.

\bibitem[Smith et~al.(2009)Smith, McIntosh, Donnelly, Funk, Mahoney, and Verghese]{10.1117/12.819126}
G.~M. Smith, K.~A. McIntosh, J.~P. Donnelly, J.~E. Funk, L.~J. Mahoney, and S.~Verghese, ``{Reliable InP-based Geiger-mode avalanche photodiode arrays},'' in \emph{Advanced Photon Counting Techniques III}, M.~A. Itzler and J.~C. Campbell, Eds., vol. 7320, International Society for Optics and Photonics.\hskip 1em plus 0.5em minus 0.4em\relax SPIE, 2009, p. 73200R.

\bibitem[Sicre et~al.(2022{\natexlab{b}})Sicre, Federspiel, Roy, Mamby, Coutier, and Calmon]{SICRE3}
M.~Sicre, X.~Federspiel, D.~Roy, B.~Mamby, C.~Coutier, and F.~Calmon, ``{Identification of stress factors and degradation mechanisms inducing DCR drift in SPADs},'' in \emph{2022 IEEE International Integrated Reliability Workshop (IIRW)}, 2022, pp. 1--5.

\bibitem[Sicre et~al.(2023{\natexlab{a}})Sicre, Roy, and Calmon]{SICREESS}
M.~Sicre, D.~Roy, and F.~Calmon, ``{Hot-Carrier Degradation modeling of DCR drift in SPADs},'' in \emph{ESSDERC 2023 - IEEE 53rd European Solid-State Device Research Conference (ESSDERC)}, 2023, pp. 61--64.

\bibitem[Lyding et~al.(1996)Lyding, Hess, and Kizilyalli]{doi:10.1063/1.116172}
J.~W. Lyding, K.~Hess, and I.~C. Kizilyalli, ``{Reduction of hot electron degradation in metal oxide semiconductor transistors by deuterium processing},'' \emph{Applied Physics Letters}, vol.~68, no.~18, pp. 2526--2528, 1996.

\bibitem[Tyaginov(2015)]{TUW-237415}
S.~E. Tyaginov, ``{P}hysics-{B}ased {M}odeling of {H}ot-{C}arrier {D}egradation,'' in \emph{{H}ot {C}arrier {D}egradation in {S}emiconductor {D}evices}, T.~Grasser, Ed.\hskip 1em plus 0.5em minus 0.4em\relax {S}pringer {I}nternational {P}ublishing, 2015, pp. 105--150.

\bibitem[McMahon et~al.(2003)McMahon, Haggag, and Hess]{1186774}
W.~McMahon, A.~Haggag, and K.~Hess, ``{Reliability scaling issues for nanoscale devices},'' \emph{IEEE Transactions on Nanotechnology}, vol.~2, no.~1, pp. 33--38, 2003.

\bibitem[Mcmahon et~al.(2002)Mcmahon, Matsuda, Lee, Hess, and Lyding]{a8e887eaa8c94af19390be1498d29f37}
W.~Mcmahon, K.~Matsuda, J.~Lee, K.~Hess, and J.~Lyding, ``{The effects of a multiple carrier model of interface trap generation on lifetime extraction for MOSFETs},'' \emph{2002 International Conference on Modeling and Simulation of Microsystems - MSM 2002}, pp. 576--579, 01 2002.

\bibitem[Hess et~al.(1998)Hess, Register, Tuttle, Lyding, and Kizilyalli]{HESS19981}
K.~Hess, L.~Register, B.~Tuttle, J.~Lyding, and I.~Kizilyalli, ``{Impact of nanostructure research on conventional solid-state electronics: The giant isotope effect in hydrogen desorption and CMOS lifetime},'' \emph{Physica E: Low-dimensional Systems and Nanostructures}, vol.~3, no.~1, pp. 1--7, 1998.

\bibitem[Hong et~al.(1999)Hong, Nam, Yun, Lee, Yu, and Park]{HONG1999809}
S.~H. Hong, S.~M. Nam, B.~O. Yun, B.~J. Lee, C.~G. Yu, and J.~T. Park, ``{Temperature dependence of hot carrier induced MOSFET degradation at low gate bias},'' \emph{Microelectronics Reliability}, vol.~39, no.~6, pp. 809--814, 1999, european Symposium on Reliability of Electron Devices, Failure Physics and Analysis.

\bibitem[Childs and Leung(1995)]{Childs1}
P.~Childs and C.~Leung, ``{New mechanism of hot carrier generation in very short channel MOSFETs},'' \emph{Electronics Letters}, vol.~31, pp. 139--141(2), January 1995.

\bibitem[Rauch et~al.(2001)Rauch, Rosa, and Guarin]{Rauch2}
S.~Rauch, G.~Rosa, and F.~Guarin, ``{Role of E-E scattering in the enhancement of channel hot carrier degradation of deep sub-micron NMOSFETs at high VGS conditions},'' vol.~1, 02 2001, pp. 399 -- 405.

\bibitem[Childs and Leung(1996)]{Childs1996AOS}
P.~A. Childs and C.~C.~C. Leung, ``{A one-dimensional solution of the Boltzmann transport equation including electron–electron interactions},'' \emph{Journal of Applied Physics}, vol.~79, no.~1, pp. 222--227, 01 1996.

\bibitem[Rauch et~al.(1998)Rauch, Guarin, and LaRosa]{Rauch1998ImpactOE}
S.~Rauch, F.~Guarin, and G.~LaRosa, ``Impact of e-e scattering to the hot carrier degradation of deep submicron nmosfets,'' \emph{IEEE Electron Device Letters}, vol.~19, no.~12, pp. 463--465, 1998.

\bibitem[Hu(1979)]{Hu1979LuckyelectronMO}
C.~Hu, ``{Lucky-electron model of channel hot electron emission},'' in \emph{1979 International Electron Devices Meeting}, 1979, pp. 22--25.

\bibitem[Bravaix et~al.(2009)Bravaix, Guerin, Huard, Roy, Roux, and Vincent]{5173308}
A.~Bravaix, C.~Guerin, V.~Huard, D.~Roy, J.~Roux, and E.~Vincent, ``{Hot-Carrier acceleration factors for low power management in DC-AC stressed 40nm NMOS node at high temperature},'' in \emph{2009 IEEE International Reliability Physics Symposium}, 2009, pp. 531--548.

\bibitem[Bravaix et~al.(2015)Bravaix, Huard, Cacho, Federspiel, and Roy]{Bravaix2}
A.~Bravaix, V.~Huard, F.~Cacho, X.~Federspiel, and D.~Roy, ``{Hot-Carrier Degradation in Decananometer CMOS Nodes: From an Energy-Driven to a Unified Current Degradation Modeling by a Multiple-Carrier Degradation Process},'' pp. 57--103, 10 2015.

\bibitem[Guerin et~al.(2009)Guerin, Huard, and Bravaix]{Guerin2}
C.~Guerin, V.~Huard, and A.~Bravaix, ``{General framework about defect creation at the Si/SiO2 interface},'' \emph{Journal of Applied Physics}, vol. 105, pp. 114\,513 -- 114\,513, 07 2009.

\bibitem[Penzin et~al.(2003)Penzin, Haggag, McMahon, Lyumkis, and Hess]{1213815}
O.~Penzin, A.~Haggag, W.~McMahon, E.~Lyumkis, and K.~Hess, ``{MOSFET degradation kinetics and its simulation},'' \emph{IEEE Transactions on Electron Devices}, vol.~50, no.~6, pp. 1445--1450, 2003.

\bibitem[Tyaginov et~al.(2010)Tyaginov, Starkov, Triebl, Cervenka, Jungemann, Carniello, Park, Enichlmair, Karner, Kernstock, Seebacher, Minixhofer, Ceric, and Grasser]{TYAGINOV20101267}
S.~Tyaginov, I.~Starkov, O.~Triebl, J.~Cervenka, C.~Jungemann, S.~Carniello, J.~Park, H.~Enichlmair, M.~Karner, C.~Kernstock, E.~Seebacher, R.~Minixhofer, H.~Ceric, and T.~Grasser, ``{Interface traps density-of-states as a vital component for hot-carrier degradation modeling},'' \emph{Microelectronics Reliability}, vol.~50, no.~9, pp. 1267--1272, 2010, 21st European Symposium on the Reliability of Electron Devices, Failure Physics and Analysis.

\bibitem[Hofbauer et~al.(2018)Hofbauer, Steindl, and Zimmermann]{Hofbauer}
M.~Hofbauer, B.~Steindl, and H.~Zimmermann, ``{Temperature Dependence of Dark Count Rate and After Pulsing of a Single-Photon Avalanche Diode with an Integrated Active Quenching Circuit in 0.35 $\mu$m CMOS},'' \emph{Journal of Sensors}, vol. 2018, pp. 1--7, 07 2018.

\bibitem[Moussy and Ouvrier-Buffet(U.S. Patent 9,780,247, B2, issued October 3, 2017)]{Moussy}
N.~Moussy and J.~Ouvrier-Buffet, ``{SPAD-TYPE PHOTODIODE},'' U.S. Patent 9,780,247, B2, issued October 3, 2017.

\bibitem[Wuu et~al.(2022)Wuu, Chen, Chien, Enquist, Guidash, and McCarten]{9732895}
S.-G. Wuu, H.-L. Chen, H.-C. Chien, P.~Enquist, R.~M. Guidash, and J.~McCarten, ``{A Review of 3-Dimensional Wafer Level Stacked Backside Illuminated CMOS Image Sensor Process Technologies},'' \emph{IEEE Transactions on Electron Devices}, vol.~69, no.~6, pp. 2766--2778, 2022.

\bibitem[Pellegrini et~al.(2017)Pellegrini, Rae, Pingault, Golanski, Jouan, Lapeyre, and Mamdy]{Pellegrini}
S.~Pellegrini, B.~Rae, A.~Pingault, D.~Golanski, S.~Jouan, C.~Lapeyre, and B.~Mamdy, ``Industrialised spad in 40 nm technology,'' 12 2017, pp. 16.5.1--16.5.4.

\bibitem[Cova et~al.(1996)Cova, Ghioni, Lacaita, Samori, and Zappa]{Cova:96}
S.~Cova, M.~Ghioni, A.~Lacaita, C.~Samori, and F.~Zappa, ``Avalanche photodiodes and quenching circuits for single-photon detection,'' \emph{Appl. Opt.}, vol.~35, no.~12, pp. 1956--1976, Apr 1996.

\bibitem[Tisa et~al.(2008)Tisa, Guerrieri, and Zappa]{Tisa2008VariableloadQC}
S.~Tisa, F.~Guerrieri, and F.~Zappa, ``Variable-load quenching circuit for single-photon avalanche diodes.'' \emph{Optics express}, vol. 16 3, pp. 2232--44, 2008.

\bibitem[spr(2019)]{sprocess}
``{Sentaurus Process User Guide},'' \emph{Version P-2019.03}, March 2019.

\bibitem[sde(2019)]{sdevice}
``{Sentaurus Device User Guide},'' \emph{Version P-2019.03}, March 2019.

\bibitem[Chynoweth(1958)]{PhysRev.109.1537}
A.~G. Chynoweth, ``{Ionization Rates for Electrons and Holes in Silicon},'' \emph{Phys. Rev.}, vol. 109, pp. 1537--1540, Mar 1958.

\bibitem[Helleboid et~al.(2022)Helleboid, Rideau, Nicholson, Grebot, Mamdy, Mugny, Basset, Agnew, Golanski, Pellegrini, Saint-Martin, Pala, and Dollfus]{Helleboid_2022}
R.~Helleboid, D.~Rideau, I.~Nicholson, J.~Grebot, B.~Mamdy, G.~Mugny, M.~Basset, M.~Agnew, D.~Golanski, S.~Pellegrini, J.~Saint-Martin, M.~Pala, and P.~Dollfus, ``A fokker–planck-based monte carlo method for electronic transport and avalanche simulation in single-photon avalanche diodes,'' \emph{Journal of Physics D: Applied Physics}, vol.~55, no.~50, p. 505102, oct 2022.

\bibitem[Jacoboni and Lugli(1989)]{Jacoboni_2}
C.~Jacoboni and P.~Lugli, \emph{{The Monte Carlo Method for Semiconductor Device Simulation}}.\hskip 1em plus 0.5em minus 0.4em\relax Computational Microelectronics, Wien: Springer-Verlag, 1989.

\bibitem[Mars(1972)]{PMARS}
P.~Mars, ``{Temperature dependence of avalanche breakdown voltage Temperature dependence of avalanche breakdown voltage in p—n junctions†},'' \emph{International Journal of Electronics}, vol.~32, no.~1, pp. 23--37, 1972.

\bibitem[Ingargiola et~al.(2009)Ingargiola, Assanelli, Gallivanoni, Rech, Ghioni, and Cova]{Ingargiola}
A.~Ingargiola, M.~Assanelli, A.~Gallivanoni, I.~Rech, M.~Ghioni, and S.~Cova, ``{Avalanche buildup and propagation effects on photon-timing jitter in Si-SPAD with non-uniform electric field},'' \emph{Proc SPIE}, vol. 7320, 05 2009.

\bibitem[Michaillat et~al.(2010)Michaillat, Rideau, Aniel, Tavernier, and Jaouen]{MICHAILLAT20102437}
M.~Michaillat, D.~Rideau, F.~Aniel, C.~Tavernier, and H.~Jaouen, ``{Full-Band Monte Carlo investigation of hole mobilities in SiGe, SiC and SiGeC alloys},'' \emph{Thin Solid Films}, vol. 518, no.~9, pp. 2437--2441, 2010, proceedings of the EMRS 2009 Spring Meeting Symposium I: Silicon and germanium issues for future CMOS devices.

\bibitem[Jacoboni and Reggiani(1983)]{RevModPhys.55.645}
C.~Jacoboni and L.~Reggiani, ``{The Monte Carlo method for the solution of charge transport in semiconductors with applications to covalent materials},'' \emph{Rev. Mod. Phys.}, vol.~55, pp. 645--705, Jul 1983.

\bibitem[Niquet et~al.(2009)Niquet, Rideau, Tavernier, Jaouen, and Blase]{Niquet}
Y.-M. Niquet, D.~Rideau, C.~Tavernier, H.~Jaouen, and X.~Blase, ``Onsite matrix elements of the tight-binding hamiltonian of a strained crystal: Application to silicon, germanium, and their alloys,'' \emph{Physical Review B}, vol.~79, p. 245201, 06 2009.

\bibitem[Grasser et~al.(2004)Grasser, Jungemann, Kosina, Meinerzhagen, and Selberherr]{GrasserJungemann}
T.~Grasser, C.~Jungemann, H.~Kosina, B.~Meinerzhagen, and S.~Selberherr, ``Advanced transport models for sub-micrometer devices,'' in \emph{Simulation of Semiconductor Processes and Devices 2004}, G.~Wachutka and G.~Schrag, Eds.\hskip 1em plus 0.5em minus 0.4em\relax Vienna: Springer Vienna, 2004, pp. 1--8.

\bibitem[Zaka et~al.(2010)Zaka, Rafhay, Iellina, Palestri, Clerc, Rideau, Garetto, Dornel, Singer, Pananakakis, Tavernier, and Jaouen]{ZakaRafhay}
A.~Zaka, Q.~Rafhay, M.~Iellina, P.~Palestri, R.~Clerc, D.~Rideau, D.~Garetto, E.~Dornel, J.~Singer, G.~Pananakakis, C.~Tavernier, and H.~Jaouen, ``On the accuracy of current tcad hot carrier injection models in nanoscale devices,'' \emph{Solid-State Electronics}, vol.~54, pp. 1669--1674, 12 2010.

\bibitem[Keldysh(1965)]{Keldysh1965ConcerningTT}
L.~V. Keldysh, ``{Concerning the Theory of Impact Ionization in Semiconductors},'' 1965.

\bibitem[Kamakura et~al.(2016)Kamakura, Fujita, Konaga, Ueoka, Mori, and Kotani]{7605145}
Y.~Kamakura, R.~Fujita, K.~Konaga, Y.~Ueoka, N.~Mori, and T.~Kotani, ``{Full band Monte Carlo simulation of impact ionization in wide bandgap semiconductors based on ab initio calculation},'' in \emph{2016 International Conference on Simulation of Semiconductor Processes and Devices (SISPAD)}, 2016, pp. 47--51.

\bibitem[Kunikiyo et~al.(1996)Kunikiyo, Takenaka, Morifuji, Taniguchi, and Hamaguchi]{10.1063/1.362375}
T.~Kunikiyo, M.~Takenaka, M.~Morifuji, K.~Taniguchi, and C.~Hamaguchi, ``{A model of impact ionization due to the primary hole in silicon for a full band Monte Carlo simulation},'' \emph{Journal of Applied Physics}, vol.~79, no.~10, pp. 7718--7725, 05 1996.

\bibitem[Fischetti and Laux(1988)]{PhysRevB.38.9721}
M.~V. Fischetti and S.~E. Laux, ``{Monte carlo analysis of electron transport in small semiconductor devices including band-structure and space-charge effects},'' \emph{Phys. Rev. B}, vol.~38, pp. 9721--9745, Nov 1988.

\bibitem[Yamada and Ferry(1995)]{YAMADA1995881}
T.~Yamada and D.~Ferry, ``{Monte Carlo simulation of hole transport in strained Si$_{1−x}$ Ge$_x$},'' \emph{Solid-State Electronics}, vol.~38, no.~4, pp. 881--890, 1995.

\bibitem[Tuttle and Van~de Walle(1999)]{PhysRevB.59.12884}
B.~Tuttle and C.~G. Van~de Walle, ``{Structure, energetics, and vibrational properties of Si-H bond dissociation in silicon},'' \emph{Phys. Rev. B}, vol.~59, pp. 12\,884--12\,889, May 1999.

\bibitem[Jech et~al.(2019)Jech, El-Sayed, Tyaginov, Shluger, and Grasser]{PhysRevB.100.195302}
M.~Jech, A.-M. El-Sayed, S.~Tyaginov, A.~L. Shluger, and T.~Grasser, ``$ab$ $initio$ treatment of silicon-hydrogen bond rupture at ${\mathrm{si}/\mathrm{sio}}_{2}$ interfaces,'' \emph{Phys. Rev. B}, vol. 100, p. 195302, Nov 2019.

\bibitem[{Van Overstraeten} and {De Man}(1970)]{VANOVERSTRAETEN1970583}
R.~{Van Overstraeten} and H.~{De Man}, ``{Measurement of the ionization rates in diffused silicon p-n junctions},'' \emph{Solid-State Electronics}, vol.~13, no.~5, pp. 583--608, 1970.

\bibitem[Nguyen et~al.(2011)Nguyen, Keunen, Afanas’ev, and Stesmans]{Nguyen}
T.~Nguyen, K.~Keunen, V.~Afanas’ev, and A.~Stesmans, ``{Interface state energy distribution and Pb defects at Si(110)/SiO2 interfaces: Comparison to (111) and (100) silicon orientations},'' \emph{Journal of Applied Physics}, vol. 109, pp. 013\,710--013\,710, 01 2011.

\bibitem[Zeghbroeck(2004)]{Zeghbroeck}
B.~V. Zeghbroeck, \emph{Principles of Semiconductor Devices}, 2004.

\bibitem[Varshni(1967)]{VARSHNI1967149}
Y.~Varshni, ``{Temperature dependence of the energy gap in semiconductors},'' \emph{Physica}, vol.~34, no.~1, pp. 149--154, 1967.

\bibitem[Pankove and Kiewit(1972)]{Pankove_1972}
J.~I. Pankove and D.~A. Kiewit, ``{Optical Processes in Semiconductors},'' \emph{Journal of The Electrochemical Society}, vol. 119, no.~5, p. 156Ca, may 1972.

\bibitem[Lenahan and Conley(1998)]{Lenahan}
P.~M. Lenahan and J.~Conley, J.~F., ``{What can electron paramagnetic resonance tell us about the Si/SiO2 system?}'' \emph{Journal of Vacuum Science and Technology B: Microelectronics and Nanometer Structures Processing, Measurement, and Phenomena}, vol.~16, no.~4, pp. 2134--2153, 07 1998.

\bibitem[Pantelides et~al.(2000)Pantelides, Rashkeev, Buczko, Fleetwood, and Schrimpf]{903763}
S.~Pantelides, S.~Rashkeev, R.~Buczko, D.~Fleetwood, and R.~Schrimpf, ``Reactions of hydrogen with si-sio/sub 2/ interfaces,'' \emph{IEEE Transactions on Nuclear Science}, vol.~47, no.~6, pp. 2262--2268, 2000.

\bibitem[Haggag et~al.(2001{\natexlab{a}})Haggag, Mcmahon, Hess, Fischer, and Register]{Haggag}
A.~Haggag, W.~Mcmahon, K.~Hess, B.~Fischer, and L.~Register, ``{Impact of Scaling on CMOS Chip Failure Rate, and Design Rules for Hot Carrier Reliability},'' \emph{VLSI Design}, vol.~13, 01 2001.

\bibitem[Shockley and Read(1952)]{PhysRev.87.835}
W.~Shockley and W.~T. Read, ``{Statistics of the Recombinations of Holes and Electrons},'' \emph{Phys. Rev.}, vol.~87, pp. 835--842, Sep 1952.

\bibitem[Biegelsen et~al.(1985)Biegelsen, Johnson, Stutzmann, Poindexter, and Caplan]{BIEGELSEN1985879}
D.~Biegelsen, N.~Johnson, M.~Stutzmann, E.~Poindexter, and P.~Caplan, ``{Native defects at the Si/SiO2 interface-amorphous silicon revisited},'' \emph{Applications of Surface Science}, vol. 22-23, pp. 879--890, 1985.

\bibitem[Hurkx et~al.(1992)Hurkx, Klaassen, and Knuvers]{121690}
G.~Hurkx, D.~Klaassen, and M.~Knuvers, ``A new recombination model for device simulation including tunneling,'' \emph{IEEE Transactions on Electron Devices}, vol.~39, no.~2, pp. 331--338, 1992.

\bibitem[Colalongo et~al.(1997)Colalongo, Valdinoci, Baccarani, Migliorato, Tallarida, and Reita]{COLALONGO1997627}
L.~Colalongo, M.~Valdinoci, G.~Baccarani, P.~Migliorato, G.~Tallarida, and C.~Reita, ``{Numerical analysis of poly-TFTs under off conditions},'' \emph{Solid-State Electronics}, vol.~41, no.~4, pp. 627--633, 1997.

\bibitem[Swinehart(1962)]{Swinehart}
D.~F. Swinehart, ``{The Beer-Lambert Law},'' \emph{Journal of Chemical Education}, vol.~39, no.~7, p. 333, 1962.

\bibitem[Stesmans(1996{\natexlab{a}})]{doi:10.1063/1.116308}
A.~Stesmans, ``{Passivation of Pb0 and Pb1 interface defects in thermal (100) Si/SiO2 with molecular hydrogen},'' \emph{Applied Physics Letters}, vol.~68, no.~15, pp. 2076--2078, 1996.

\bibitem[Stesmans(1996{\natexlab{b}})]{Stesmans}
------, ``{Revision of H2 passivation of P2 interface defects in standard (111)Si/SiO2},'' \emph{Applied Physics Letters}, vol.~68, no.~19, p. 2723–2725, 1996.

\bibitem[Erdogan et~al.(1994)Erdogan, Mizrahi, Lemaire, and Monroe]{doi:10.1063/1.357062}
T.~Erdogan, V.~Mizrahi, P.~J. Lemaire, and D.~Monroe, ``{Decay of ultraviolet‐induced fiber Bragg gratings},'' \emph{Journal of Applied Physics}, vol.~76, no.~1, pp. 73--80, 1994.

\bibitem[Haggag et~al.(2001{\natexlab{b}})Haggag, McMahon, Hess, Cheng, Lee, and Lyding]{922913}
A.~Haggag, W.~McMahon, K.~Hess, K.~Cheng, J.~Lee, and J.~Lyding, ``{High-performance chip reliability from short-time-tests-statistical models for optical interconnect and HCI/TDDB/NBTI deep-submicron transistor failures},'' in \emph{2001 IEEE International Reliability Physics Symposium Proceedings. 39th Annual (Cat. No.00CH37167)}, 2001, pp. 271--279.

\bibitem[Bude and Hess(1992)]{doi:10.1063/1.351434}
J.~Bude and K.~Hess, ``{Thresholds of impact ionization in semiconductors},'' \emph{Journal of Applied Physics}, vol.~72, no.~8, pp. 3554--3561, 1992.

\bibitem[Blat et~al.(1991)Blat, Nicollian, and Poindexter]{Blat}
C.~E. Blat, E.~H. Nicollian, and E.~H. Poindexter, ``{Mechanism of negative‐bias‐temperature instability},'' \emph{Journal of Applied Physics}, vol.~69, no.~3, pp. 1712--1720, 1991.

\bibitem[Sicre et~al.(2023{\natexlab{b}})Sicre, Federspiel, Mamdy, Roy, and Calmon]{SICRE4}
M.~Sicre, X.~Federspiel, B.~Mamdy, D.~Roy, and F.~Calmon, ``{Characterization and modeling of DCR and DCR drift variability in SPADs},'' in \emph{2023 IEEE International Reliability Physics Symposium (IRPS)}, 2023, pp. 1--5.

\end{thebibliography}


\end{document}